\documentclass[]{jfm}

\usepackage{graphicx}
\usepackage[normalem]{ulem}
\usepackage{newtxtext}
\usepackage{newtxmath}
\usepackage{natbib}
\usepackage{hyperref}
\hypersetup{
    colorlinks = true,
    urlcolor   = blue,
    citecolor  = black,
}

\newcommand{\RomanNumeralCaps}[1]
\linenumbers

\title{Evolution of perturbed long nonlinear plane, ring, and hybrid surface waves}
\author{Benjamin Martin\aff{1}, Dmitri Tseluiko\aff{1}, Karima Khusnutdinova\aff{1}
\corresp{\email{K.Khusnutdinova@lboro.ac.uk}}}
\affiliation{
\aff{1}Department of Mathematical Sciences, Loughborough University, Loughborough LE11 3TU, UK}

\begin{document}
\maketitle

\begin{abstract}
{\color{black} The  two-dimensional evolution of perturbed long weakly-nonlinear surface plane, ring, and hybrid waves, consisting, to leading order, of a part of a ring and two tangent plane waves,   is modelled numerically within the scope of the 2D Boussinesq--Peregrine system. Numerical runs are initiated and interpreted using the reduced 2+1-dimensional cKdV-type  and KPII equations. The cKdV-type equation leads to two different models, the KdV$\theta$ and cKdV equations, depending on whether we use the general or singular  (i.e. the envelope of the general) solution of the associated nonlinear first-order differential equation. The KdV$\theta$ equation is also derived directly from the 2D Boussinesq--Peregrine system and  used to analytically describe the intermediate 2D asymptotics of line solitons subject to sufficiently long transverse perturbations of finite  strength, while the cKdV equation is used to  initiate outward- and inward-propagating ring waves with localised and periodic perturbations. Both of these equations,  together with the KPII equation, are used to model the evolution of hybrid waves, where we show, in particular, that large localised waves (lumps) can appear as transient  (emerging and then disappearing) states in the evolution of inward-propagating waves, contributing to the possible mechanisms for the generation of rogue waves. Detailed comparisons are made between the key features of the non-stationary two-dimensional modelling and relevant predictions of the reduced equations.}  
\end{abstract}

\begin{keywords}
Authors should not enter keywords on the manuscript, as these must be chosen by the author during the online submission process and will then be added during the typesetting process (see \href{https://www.cambridge.org/core/journals/journal-of-fluid-mechanics/information/list-of-keywords}{Keyword PDF} for the full list).  Other classifications will be added at the same time.
\end{keywords}

%{\bf MSC Codes }  {\it(Optional)} Please enter your MSC Codes here

% ---------- ---------- ---------- ---------- % 
\section{Introduction}

\begin{figure}
	\centerline{\includegraphics[width=0.7 \linewidth]{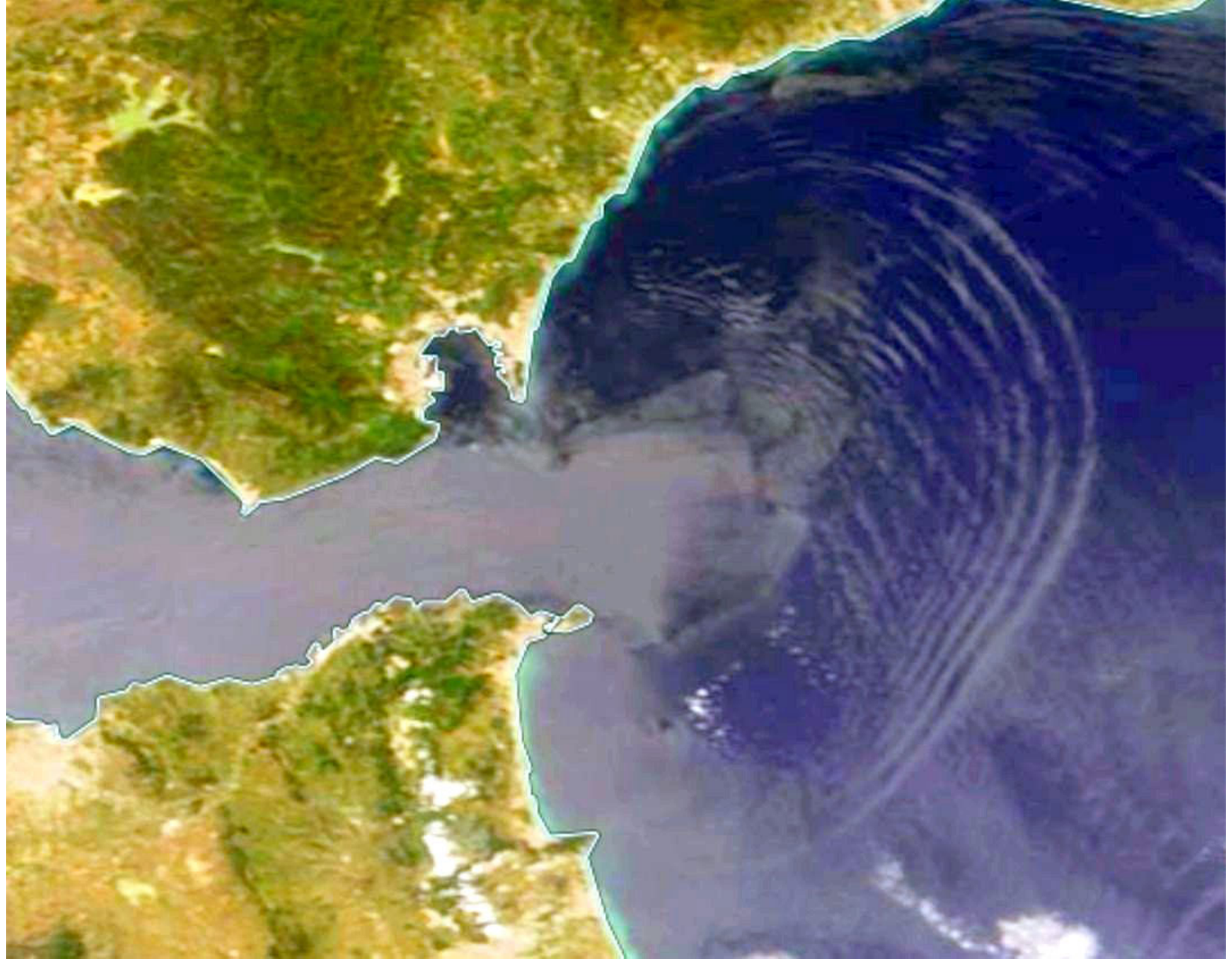}}
	\caption{Hybrid wave generated in the Strait of Gibraltar (Terra MODIS, 22 May 2017, 11:30 UTC. NASA ESDIS Worldview, https://user.eumetsat.int/resources/case-studies/internal-waves-in-the-eastern-strait-of-gibraltar). }
\label{HW}
\end{figure}

{\color{black} SAR images of the waves generated in narrow straits and river--sea interaction areas often look like a part of a ring wave with two attached tangent plane waves (see, for example, figure \ref{HW} for the image of a hybrid wave generated in the Strait of Gibraltar).  To the best of our knowledge, the first mentioning of the possibility of existence of stable outward-propagating hybrid waves in theoretical literature was due to \citet{OS}, who also predicted, using ray theory, that inward-propagating hybrid waves are unstable. However, there have been no attempts to systematically model the evolution of such 2D waves. This raises several key questions. Does numerical modelling support their existence? Is there indeed any difference between the outward- and inward-propagation? If there is some nontrivial evolution scenario in the latter case, will its key features persist under perturbations? It is also important to determine, which theoretical framework can be used to initiate the wave  and interpret the features observed in numerical runs. Hence, the primary aim of our study is the 2D modelling of the propagation and evolution of both outward- and inward-propagating hybrid waves, which, in this first study on the topic, is performed within the scope of the 2D Boussinesq--Peregrine \citep{P} system for surface waves in a homogeneous fluid. However, we also outline how theoretical considerations can be extended to the general setting of surface and internal waves in a stratified fluid on a parallel shear flow. 

In order to initiate the numerical runs and interpret their results, we use approaches based on asymptotic multiple-scale expansions describing weakly-nonlinear plane and ring waves, and their 2D extensions. Naturally, since we view the hybrid waves as a combination of the elements of plane and ring waves (at least initially), we also model the evolution of the perturbed plane and ring waves {\it per se}.  Approaching the plane waves from a new perspective allows us to derive a new amplitude equation (which we call the KdV$\theta$ equation), which is used to obtain analytical results for plane waves subject to long transverse perturbations. We also model the evolution of 2D perturbations of ring waves. Different aspects of the stability of ring waves is a topic of active current research (see \citealt{K, HKG, K24, HPS, ZHGS} and the references therein). 

The weakly-nonlinear theory for plane waves leading to the Korteweg--de Vries (KdV) equation and its extensions are an important theoretical framework which allows one to understand and interpret many features of the behaviour of both surface and internal plane waves in the ocean (see, for example,  \citealt{GOSS, G, L, HM, Gr, AOSL, BLS, GPTK, AB, GHJ, CK, OPSS, S, STCK} and references therein). Derivation of the KdV equation in the general setting of a stratified fluid with a parallel depth-dependent shear flow is based on the existence of the modal decomposition in the set of Euler equations with boundary conditions appropriate for oceanographic applications. In this decomposition, the leading-order vertical particle displacement $\zeta$ (with $w = d \zeta / dt$ defining the vertical velocity) is given by 
\begin{equation}
\zeta = A(T, \tilde \xi) \phi(z), 
\label{1}
\end{equation}
where $A(T, \tilde \xi)$ is the amplitude function satisfying the KdV equation ($T = \epsilon t, \tilde \xi = x - c_0 t$, where $c_0$ is the linear long-wave speed of plane waves propagating in the $x$-direction, $\epsilon$ is the small-amplitude parameter), and $\phi(z)$ is the modal function which should be found from the set of appropriate modal equations (for details see, for example, \citealt{G} and references therein). 

In contrast to this well developed theory, asymptotic theory describing long surface ring waves in a homogeneous fluid  propagating over a parallel depth-dependent shear flow was developed relatively recently by \citet{J90}. The theory was generalised  to describe long surface and internal ring waves in a stratified fluid with the density $\rho_0(z)$ and a parallel depth-dependent shear flow $u_0(z)$ by \citet{KZ}. The generalisation was based on finding a new far-field modal decomposition in the set of Euler equations, more complicated than that for plane waves. In particular, rather than (\ref{1}), the leading-order vertical particle displacement for the ring waves is given by 
\begin{equation}
\zeta = A(R, \xi, \theta) \phi(z, \theta), 
\end{equation}
where $r, \theta$ are polar coordinates in the reference frame moving at a constant speed $c$, $R = \epsilon rk(\theta)$ is a slow variable, $\xi = rk(\theta) - st$ is a fast variable,  $s$ is the wave speed in the absence of any shear flow, $k(\theta)$ is a function accounting for the effect of a shear flow on the linear long-wave speed in different directions, and $\epsilon$ is again the small-amplitude parameter. In non-dimensional variables, the bottom is at $z = 0$, the unperturbed free surface is at $z = 1$.

For the given density $\rho_0(z)$ and shear flow $u_0(z)$, the modal function $\phi(z, \theta)$, the function $k(\theta)$ and admissible wave speeds $s$ should be found by solving the following boundary-value problem (modal equations):
\begin{eqnarray}
&&\left (\frac{\rho_0 F^2}{k^2 + k^{'2}} \phi_z\right )_z -  \rho_{0z} \phi = 0 \quad \mbox{for} \quad  0 < z < 1, \label{modal1} \\
&&\frac{F^2}{k^2 + k^{'2}} \phi_z -  \phi = 0 \quad \mbox{at} \quad z = 1, \label{modal2} \\
&&\phi = 0 \quad \mbox{at} \quad z = 0, \label{modal3} \\
&&\mbox{where} \quad k' = \frac{\mathrm{d}k}{\mathrm{d}\theta} \quad \mbox{and} \quad F = -s + (u_0 - c) (k \cos \theta - k' \sin \theta),
\end{eqnarray}
and the natural choice of the constant $c$ is the speed of the flow at the bottom, i.e.~$c = u_0(0)$.

The amplitude equation has the form
\begin{equation}
\mu_1 A_R + \mu_2 A A_{\xi} + \mu_3 A_{\xi \xi \xi} + \mu_4 \frac{A}{R} + \mu_5 \frac{A_\theta}{R} = 0, \label{A}
\end{equation}
where the coefficients are given by
\begin{eqnarray}
&&\mu_1 = 2 s \int_0^1 \rho_0 F \phi_z^2 ~ \mathrm{d}z, \
% \label{c1} \
\mu_2 = - 3 \int_0^1 \rho_0 F^2 \phi_z^3 ~ \mathrm{d}z,  \
%\label{c2} 
\mu_3 = - (k^2 + k'^2) \int_0^1  \rho_0 F^2 \phi^2 ~ \mathrm{d}z, \qquad \label{c3} 
\end{eqnarray}
\begin{eqnarray}
&&\mu_4 = - \int_0^1 \left \{  \frac{\rho_0 \phi_z^2 k (k+k'')}{(k^2+k'^2)^2} \left [ (k^2-3k'^2) F^2 -
{4 k' (k^2 + k'^2) (u_0-c)\sin \theta} F  \right .  \right . \nonumber \\
&&\left . \left .  - \sin^2 \theta (u_0-c)^2(k^2 + k'^2)^2 \right ] + \frac{2 \rho_0 k}{k^2 + k'^2} F \phi_z \phi_{z\theta} [k' F + (k^2 + k'^2) (u_0-c) \sin \theta ] \right \} \mathrm{d}z, \qquad \label{c4} \\
&& \mu_5 = - \frac{2k}{k^2 + k'^2} \int_0^1 \rho_0 F \phi_z^2 [k' F + (u_0-c) (k^2+k'^2) \sin \theta ] \ \mathrm{d}z.  \label{c5}
\end{eqnarray}

For the particular case of surface waves in a homogeneous fluid with the non-dimensional density $\rho_0 = 1$, we have the following solution of the modal equation (\ref{modal1}) with the boundary condition (\ref{modal3}):
\begin{equation}
\phi = \Lambda (k^2+k'^2) \int_0^z\frac {1}{F^2}\,\mathrm{d}z, \label{2}
\end{equation}
where $\Lambda$ is a parameter. Next, assuming that $u_0(z) = 0$ (no shear flow), setting $k=1$ and $s=1$ (concentric wavefronts, non-dimensional wave speed), the boundary condition (\ref{modal2}) yields $\Lambda=1$, while in the presence of the shear flow we recover, from the same boundary condition, the “generalised Burns condition” \citep{J90}
\begin{equation}
(k^2+k'^2)\int_0^1\frac {1}{F^2}\,\mathrm{d} z = 1, \label{GB}
\end{equation}
while  the modal function is given by
\begin{equation}
\phi = (k^2+k'^2)\int_0^z\frac {1}{F^2}\,\mathrm{d}z. \label{phi}
\end{equation}
Equation (\ref{GB}) is a nonlinear first-order ordinary differential equation which admits both the general solution (one-parameter family of solutions) and the singular solution -- an additional solution, given by the envelope of the general solution (see \citealt{J90} and references therein). The general solution of (\ref{GB}) is dependent on a single constant parameter $a$ and  has the form 
\begin{equation}
k(\theta; a) = a \cos \theta + b(a) \sin \theta, \quad \mbox{where} \quad  [a^2 + b(a)^2] \int_0^1 \frac{\mathrm{d}z}{[1-(u_0(z)-c) a]^2} = 1. \label{gs}
\end{equation}
The singular solution is then given by the geometrical envelope of the family of curves described by the general solution.  It can be found by eliminating the parameter $a$ from the function $k(\theta; a)$  using the condition $\displaystyle  \mathrm{d}k / \mathrm{d}a = 0$, which defines the function $a = a(\theta)$ and yields the following singular solution:
\begin{eqnarray}
&&k(\theta) = a \cos \theta + b(a) \sin \theta,  \quad \mbox{where} 
\qquad  \label{ss1}\\
&&[a^2 + b(a)^2] \int_0^1 \frac{\mathrm{d}z}{[1-(u_0(z)-c) a]^2} = 1  \quad \mbox{and} \quad
\frac{\mathrm{d}b}{\mathrm{d}a} = - \frac{1}{\tan \theta}. \qquad  \label{ss2}
\end{eqnarray}
%The latter equation follows from the condition $\displaystyle  \mathrm{d}k / \mathrm{d}a = 0$, which defines the function $a = a(\theta)$.  
There are no parameters in the singular solution, and the associated waves are ring waves \citep{J90}. We also note that the general solution  (\ref{gs}) satisfies the equation 
\begin{equation}
k'' + k = 0, \label{3} 
\end{equation}
while the singular solution cannot satisfy this equation since it is not a part of the general solution.

The equation derived by \citet{J90} has the same form (\ref{A}), but the coefficients are given by rather complicated formulae involving multiple integrals. The new formulation (\ref{c3})--(\ref{c5})   allows it to be proved that, for surface ring waves in a homogeneous fluid, the coefficient $\mu_5 = 0$ for any shear flow $u_0(z)$  \citep{KZ}. Indeed, differentiating the generalised Burns condition (\ref{GB}) with respect to $\theta$, we obtain:
\begin{equation}
2 (k+k'')  \int_0^1\frac{k'F+(u_0-c)(k^2+k'^2)\sin\theta}{F^3} \,\mathrm{d} z = 0,
\label{3a}
\end{equation}
where $k+k''\ne 0$ on the singular solution, implying that the integral in (\ref{3a}) is equal to zero instead. Next, substituting  function (\ref{phi}) into equation  (\ref{c5}), one obtains
\begin{equation}
\mu_5 = - 2\rho_0 k (k^2+k'^2)  \int_0^1\frac{k'F+(u_0-c)(k^2+k'^2)\sin\theta}{F^3} \,\mathrm{d}  z,
\label{4a}
\end{equation}
implying $\mu_5 = 0$ since the integral in (\ref{4a}) is the same as in (\ref{3a}).
Hence, long surface ring waves in a homogeneous fluid are actually described by a simpler 1+1-dimensional cKdV-type equation
\begin{equation}
\mu_1 A_R + \mu_2 AA_{\xi} + \mu_3 A_{\xi\xi\xi} + \mu_4 \frac{A}{R}  = 0. \label{cKdV}
\end{equation}

Among some other curious features revealed using the formulation based on the far-field modal decomposition was the qualitatively different behaviour of the wavefronts of surface and interfacial waves propagating over the same flows. For example,  the wavefront of the surface ring wave in a two-layer fluid with a piecewise-constant current of increasing strength approaching a critical value was elongated in the direction of the flow, while the wavefront of the interfacial wave was squeezed. This phenomenon was linked to the presence of long-wave instability  of plane waves tangent to the ring wave and propagating in the downstream and upstream directions for sufficiently strong currents (see \citealt{KZ} and  \citealt{O, BM, BC, LM}). Similar effects were shown to take place in a two-layer fluid with a two-parameter family of upper-layer currents $\displaystyle u_0 = \gamma \left ( (z+d)/d \right )^\alpha$ for sufficiently small values of $\alpha$ (with $-d$ denoting the position of the interface) by \cite{K, HKG} and in a three-layer fluid for a current with a constant vertical shear (for the second interfacial mode) by \cite{TABK}.

Interestingly, while ring waves associated with the singular solution of the equation for the function $k(\theta)$ (for example, the generalised Burns condition (\ref{GB}))  have been studied, waves associated with the general solution were overlooked. Their role in describing plane waves tangent to the ring wave was understood only recently, in connection with the kinematic consideration of hybrid wavefronts on depth-dependent parallel shear flows  \citep{HKG}, while the related amplitude equation has not been considered at all.  One of the aims of the present study is to derive this new amplitude equation
\begin{equation}
\mu_1 A_R + \mu_2 A A_{\xi} + \mu_3 A_{\xi \xi \xi}  + \mu_5 \frac{A_\theta}{R} = 0, \label{A2}
\end{equation}
which we call the KdV$\theta$ equation, and understand its role. The overarching aim of the paper, however, is to study the evolution of perturbed nonlinear plane, ring and hybrid waves within the scope of the full 2D Boussinesq--Peregrine system for surface waves \citep{P} and to identify features which can be described using the reduced amplitude equations. 

In this paper we will restrict our numerical considerations to the case of long surface waves in a homogeneous fluid without a shear flow, although all constructions have a counterpart in stratified fluids and with an underlying shear flow, as discussed above. In Section 2 we discuss the 2D Boussinesq--Peregrine system and derive the 2D counterparts of the approximate conservation laws considered, in the 1D setting, by  \citet{KMS,KK} (see Appendix \ref{appA} for the details).  We also derive a particular version of the KdV$\theta$ equation (\ref{KdVTHETA}) directly from this parent system, and include a brief derivation of the known cKdV and KPII equations (see, for example, \citealt{M2, AS, J97, STCK} and further references therein). The numerical method used to solve the system is discussed in Appendix \ref{appB}. 

In Section 3 we derive the KdV$\theta$ equation (\ref{A2}) from the 2+1-dimensional cKdV-type equation (\ref{A}) for the general setting of a stratified fluid over a parallel shear flow.  Then, we show that the KdV$\theta$ equation can be mapped to the KdV equation, but the initial condition retains the parametric dependence on the polar angle $\theta$. This allows us to make analytical predictions concerning the 2D fission of a perturbed line soliton, using the Inverse Scattering Transform \citep{GGKM}. Long transverse perturbations of line solitons modelled by the Kadomstev--Petviashvili (KP) equation (in particular, KPII equation, describing water waves) were extensively studied, both from the applied and rigorous perspectives (see \citealt{KP, P91, KP00, MT, Mi, MS1, HP, MS2} and references therein). However, to the best of our knowledge, the amplitude equation discussed in our paper has not been derived or used before. We show that it can be used instead of the KPII equation to describe the intermediate evolution of sufficiently long perturbations of finite amplitude. The reduction to the KdV equation is discussed in Appendix \ref{appC}.

Next, in Sections 4 and 5, we consider perturbed ring and hybrid waves. Our simulations within the full parent system allow us to investigate the long-time evolution subject to both localised and periodic perturbations. In particular, we find examples illustrating the difference in the behaviour of inward-propagating ring waves depending on the wavelength of the perturbation, which was predicted theoretically by \citet{OS, SP, P91}. We show that the evolution of the hybrid wave is strongly dependent on whether it is a diverging or converging wave, and that the outward-propagating (diverging) wave is stable. However, the inward-propagating (converging) case leads to a very peculiar instability scenario: there appears a large rogue wave (lump) which disintegrates into several pieces and eventually evolves into a stable ``X-type'' structure known from the theory of the KPII equation (e.g. \citealt{AC,CK,YW}). Finally, in Section 6 we summarise our results and make remarks about some natural extensions of our present studies.}

% ---------- ---------- ---------- ---------- %
\section{{\color{black} The 2D Boussinesq--Peregrine system and reduced models}}

We consider surface waves in a homogeneous fluid within the scope of the non-dimensional 2D Boussinesq--Peregrine system \citep{P}, {\color{black} assuming a flat bottom:}
\begin{eqnarray}
&&\eta_t + \nabla \cdot \{(1 + \epsilon \eta) \mathbf{u} \} = 0,  \label{Bouss1} \\
&&\mathbf{u}_t + \epsilon (\mathbf{u} \cdot \nabla ) \mathbf{u} + \nabla \eta - \frac{\delta^2}{3} \nabla (\nabla \cdot \mathbf{u}_t ) = \mathbf{0}, \label{Bouss2}
\end{eqnarray}
where $\mathbf{u} = (u,v)^{\text{T}}$ are the depth averaged horizontal velocities, $\eta$ is the free surface elevation, $\epsilon$ is the small-amplitude parameter, and $\delta$ is the long-wavelength parameter. 

{\color{black} The 2D Boussinesq--Peregrine system conserves mass exactly, while momentum and energy are conserved asymptotically (the details of our calculations can be found in Appendix \ref{appA}).} The integral forms for these conservation laws can be written as
\begin{eqnarray}
\frac{\mathrm{d} \mathcal{M}}{\mathrm{d} t} &&= \frac{\mathrm{d}}{\mathrm{d} t} \iint_D  \eta ~ \mathrm{d}x \mathrm{d}y = 0, 
\end{eqnarray}
\begin{eqnarray}
\frac{\mathrm{d} \mathcal{P}}{\mathrm{d} t} &&= \frac{\mathrm{d}}{\mathrm{d} t} \iint_D  \left( 1 + \epsilon \eta \right) \mathbf{u} ~ \mathrm{d}x \mathrm{d}y = \textit{O}(\epsilon^2,\epsilon\delta^2,\delta^4), \\
\frac{\mathrm{d} \mathcal{E}}{\mathrm{d} t} &&=  \frac{\mathrm{d}}{\mathrm{d} t} \iint_D  \frac{1}{2} \left( |\mathbf{u}|^2 + \eta^2 \right) ~ \mathrm{d}x \mathrm{d}y = \textit{O}(\epsilon,\delta^2),
\end{eqnarray}
respectively, where $D= \{(x, y)| x_1 \le x \le x_2, y_1 \le y \le y_2\}$ is the doubly periodic rectangular computational domain. {\color{black} These results generalise the 1D results obtained by \citet{KMS,KK}.} In order to match the derivation of the reduced amplitude equations we choose the balance $\epsilon = \delta^2$. {\color{black} We therefore expect mass to be conserved  in numerical simulations to machine precision, while the deviations  in momentum and energy should scale as $\epsilon^2$ and $\epsilon$, respectively.   This was indeed the case in our simulations (see the discussion in Appendix \ref{appA}).  We numerically solve the system (\ref{Bouss1})--(\ref{Bouss2}) using the pseudospectral method outlined in Appendix \ref{appB}.}

{\color{black} Next, we derive the KdV$\theta$ and cKdV equations directly from the system (\ref{Bouss1})--(\ref{Bouss2}), in the slow-radius form. } We apply the non-axisymmetric change of variables $(t,x,y) \rightarrow (t,r,\theta)$, to give
\begin{eqnarray}
&&\eta_t + U_r + \frac{U}{r} + \frac{V_{\theta}}{r} + \epsilon \left[U\eta_r + \eta U_r + \frac{U\eta + V\eta_{\theta} + \eta V_{\theta}}{r} \right] = 0, \label{trt1} \\
&&U_t + \eta_r + \frac{\epsilon}{3} \left[\frac{U_t + V_{t\theta}}{r^2} - \frac{U_{rt} - 3VU_{\theta} + 3V^2 - V_{rt\theta}}{r} - U_{rrt} + 3UU_r \right] = 0, \label{trt2} \\
&&V_t + \frac{\eta_{\theta}}{r} + \frac{\epsilon}{3}\left[-\frac{U_{t\theta} + V_{t\theta\theta}}{r^2} - \frac{3UV + 3VV_{\theta} - U_{tr\theta}}{r} + 3UV_r \right] = 0, \label{trt3}
\end{eqnarray}
where the variables $U$ and $V$ are the cylindrically projected velocity components of $\mathbf{u}$:
\begin{eqnarray}
u = U \cos\theta - V \sin\theta, \label{cartu} \quad v = U \sin\theta + V \cos\theta. \label{cartv}
\end{eqnarray}
To leading order, we obtain
\begin{eqnarray}
&&  \eta_{tt} - \eta_{rr} - \frac{\eta_r}{r} - \frac{\eta_{\theta\theta}}{r^2} = 0, \label{Lo1} \\
&&  U_t + \eta_r = 0, \label{Lo2} \\
&&  V_t + \frac{\eta_{\theta}}{r} = 0,
\end{eqnarray}
where $\eta = \eta(\xi)$, given $\xi = rk(\theta) - t$, is an exact solution of (\ref{Lo1}) if $k(\theta)$ {\color{black} has the form }  $k(\theta) = a\cos\theta + b(a)\sin\theta$ and $a^2 + b(a)^2 = 1$ {\color{black} (coinciding with the general solution (\ref{gs}) for $u_0(z) = 0$)}, e.g. $a = \cos \varphi$ and $b = \sin \varphi$, where $\varphi$ then has the meaning of the angle between the normal to the wavefront and the positive $x$-direction.
{\color{black} In particular, if $\varphi = 0$, i.e. waves propagate in the $x$-direction, then $\xi = r \cos \theta - t = x - t$.}

We then consider solutions in the far field such that $R = \epsilon rk(\theta) \sim \textit{O}(1)$ and apply the change of variables $(t,r,\theta) \rightarrow (R,\xi,\theta)$ to obtain
\begin{eqnarray}
&&\!\!\!\!\!\eta_{\xi} - kU_{\xi} - k'V_{\xi} - \epsilon \biggl[ kU_R + k\frac{U}{R} + k'V_R 
+ k\frac{V_{\theta}}{R} + k(U\eta)_{\xi} + k'(V\eta)_{\xi} \biggr]=O(\epsilon^2), \label{rxt1} \\
&&\!\!\!\!\!U_{\xi} - k\eta_{\xi} - \frac{\epsilon}{3}\biggl[3k\eta_R + 3k'VU_{\xi} 
+ kk' V_{\xi\xi\xi} + k^2 U_{\xi\xi\xi} + 3kUU_{\xi} \biggr]=O(\epsilon^2), \label{rxt2}  \\
&&\!\!\!\!\!V_{\xi} - k'\eta_{\xi} - \frac{\epsilon}{3}\biggl[3k'\eta_R + 3k\frac{\eta_{\theta}}{R} + 3k'VV_{\xi} 
+ kk' U_{\xi\xi\xi} + 3kUV_{\xi} - k'^2 V_{\xi\xi\xi} \biggr]=O(\epsilon^2).\  \qquad \label{rxt3}
\end{eqnarray}

We seek a solution of (\ref{rxt1})--(\ref{rxt3}) in the form of an asymptotic multiple-scale expansion 
\begin{equation}
\eta(R,\xi,\theta) = \eta^{(0)}(R,\xi,\theta) + \epsilon \eta^{(1)}(R,\xi,\theta) + \textit{O}(\epsilon^2) \quad \mbox{as} \quad \epsilon \rightarrow 0,
\end{equation}
and similarly for $U$ and $V$. At leading order, $O(1)$, we obtain
\begin{eqnarray}
    \eta^{(0)} &&= \eta^{(0)}(R,\xi,\theta), \\
    U^{(0)} &&= k\eta^{(0)}, \label{LO11} \\
    V^{(0)} &&= k'\eta^{(0)}, \label{LO22}
\end{eqnarray}
assuming that any disturbances are generated only by the propagating waves. At next order, $O(\epsilon)$, we obtain
{\small
\begin{eqnarray}
 &&   \eta_{\xi}^{(1)} - kU_{\xi}^{(1)} - k'V_{\xi}^{(1)} = kU_R^{(0)} + k\frac{U^{(0)}}{R} + k'V_R^{(0)} + k\frac{V_{\theta}^{(0)}}{R}  \nonumber \\
&&\hspace{3.3cm}  + \ k(U^{(0)}\eta^{(0)})_{\xi} + k'(V^{(0)}\eta^{(0)})_{\xi}, \qquad \label{Oeps1} \\
&&U_{\xi}^{(1)} = k\eta_{\xi}^{(1)} + k\eta_R^{(0)} + k'V^{(0)}U^{(0)}_{\xi} 
 + \frac{1}{3}kk'V_{\xi\xi\xi}^{(0)} + \frac{1}{3}k^2U_{\xi\xi\xi}^{(0)} + kU^{(0)}U_{\xi}^{(0)}, \qquad \label{Oeps2} \\
&&V_{\xi}^{(1)} = k'\eta_{\xi}^{(1)} + k'\eta_R^{(0)} + k\frac{\eta_{\theta}^{(0)}}{R} + k'V^{(0)}V_{\xi}^{(0)}
  + \frac{1}{3}kk' U_{\xi\xi\xi}^{(0)} + k'^2V_{\xi\xi\xi}^{(0)} + kU^{(0)}V_{\xi}^{(0)}, \qquad \label{Oeps3}
\end{eqnarray}
}
from which substituting (\ref{LO11})--(\ref{LO22}) and (\ref{Oeps2})--(\ref{Oeps3}) into (\ref{Oeps1}), to eliminate $\eta^{(1)}$, $U^{(0)}$, $U^{(1)}$, $V^{(0)}$ and $V^{(1)}$, gives the KdV$\theta$ equation for perturbed surface plane waves propagating at an  angle $\varphi$ to the positive $x$-direction as
\begin{align}
2\eta_R^{(0)} + 3 \eta^{(0)}\eta_{\xi}^{(0)} + \frac{1}{3}\eta_{\xi\xi\xi}^{(0)} + 2kk' \frac{\eta_{\theta}^{(0)}}{R} = 0.\label{Bouss_KdVT}
\end{align}
The corresponding leading-order Cartesian velocities are calculated from (\ref{cartu})--(\ref{cartv}) and (\ref{LO11})--(\ref{LO22}) to be 
\begin{equation}
u^{(0)} = \cos\varphi ~ \eta^{(0)} \qquad \text{and} \qquad v^{(0)} = \sin\varphi ~ \eta^{(0)}. \label{kdvtvel}
\end{equation}
If $x$ is chosen to be the direction of propagation, then $\varphi = 0, \ k = \cos \theta$, and equation (\ref{Bouss_KdVT})  takes the form 
\begin{equation}
2\eta_R^{(0)} + 3 \eta^{(0)}\eta_{\xi}^{(0)} + \frac{1}{3}\eta_{\xi\xi\xi}^{(0)} -2\sin\theta\cos\theta \frac{\eta_{\theta}^{(0)}}{R} = 0,
\label{KdVT}
\end{equation}
where $ R = \epsilon x$, $\xi = x-t$ and the corresponding Cartesian velocities are $u^{(0)} = \eta^{(0)}$, and $v^{(0)} = 0$. {\color{black} This is a particular form of the KdV$\theta$ equation (\ref{A2}), derived directly from the 2D Boussinesq--Peregrine system (\ref{Bouss1})--(\ref{Bouss2}). }

The cKdV equation can be derived using a simpler asymptotic expansion {\color{black} (see, for example,  \citealt{M2, J97, STCK} and references therein)}. We apply the axisymmetric change of variables $(t,x,y) \rightarrow (t,r)$ to (\ref{Bouss1})--(\ref{Bouss2}), which yields
\begin{eqnarray}
&&\eta_t + U_r + \frac{U}{r}  + \epsilon \left[U\eta_r + \eta U_r + \frac{U\eta}{r} \right] = 0, \\
&&U_t + \eta_r + \frac{\epsilon}{3} \left[\frac{U_t}{r^2} - \frac{U_{rt}}{r} - U_{rrt} + 3UU_r \right] = 0,
\end{eqnarray}
where $U$ is the radial cylindrical velocity component of $\mathbf{u}$. To leading order, we therefore obtain
\begin{eqnarray}
&&\eta_{tt} - \eta_{rr} - \frac{\eta_r}{r} = 0, \label{cKdVLO1} \\
&&u_t + \eta_r = 0, 
\end{eqnarray}
where $\eta = \eta(\xi)$, given $\xi = r - t$, is an exact solution of (\ref{cKdVLO1}). We then perform the change of variables $(t,r) \rightarrow (R,\xi)$, where $R = \epsilon r$ is the slow radius variable and $\xi = r-t$ is the fast spacial variable in a moving coordinate frame, giving
\begin{eqnarray}
&&\eta_{\xi} - U_{\xi} - \epsilon\left[U_R + \frac{U}{R} + U\eta_{\xi} + \eta U_{\xi} \right] = O(\epsilon^2), \label{ckdvexp1} \\
&&U_{\xi} - \eta_{\xi} - \frac{\epsilon}{3}\left[3\eta_R + 3UU_{\xi} + U_{\xi\xi\xi} \right] = O(\epsilon^2). \label{ckdvexp2}
\end{eqnarray}
We now seek a solution of (\ref{ckdvexp1}) and (\ref{ckdvexp2}) in the form:
\begin{equation}
\textcolor{black}{ \eta(R,\xi) = \eta^{(0)}(R,\xi) + \epsilon \eta^{(1)}(R,\xi) + \textit{O}(\epsilon^2) \quad \mbox{as} \quad \epsilon \rightarrow 0, }
\end{equation}
and similarly for the variable $U$. We then consider solutions in the far field such that $R \sim \textit{O}(1)$, which to leading order yields
\begin{equation}
\eta_{\xi}^{(0)} - U_{\xi}^{(0)} = 0, \label{LO_eta}
\end{equation}
giving that $U^{(0)} = \eta^{(0)}$, assuming that any disturbances are generated only by the propagating waves.  {\color{black} At next order,} we find that
\begin{eqnarray}
&&\eta_{\xi}^{(1)} - U_{\xi}^{(1)} = U_R^{(0)} + \frac{U^{(0)}}{R} + 2\eta^{(0)}\eta^{(0)}_{\xi} , \label{Oeps11} \\
&&U_{\xi}^{(1)} = \eta_{\xi}^{(1)} + \eta_R^{(0)} + U^{(0)}U^{(0)}_{\xi} + \frac{1}{3}U_{\xi\xi\xi}^{(0)}, \label{Oeps22}
\end{eqnarray}
which after substituting (\ref{LO_eta}) and (\ref{Oeps22}) into (\ref{Oeps11}) gives the axisymmetric cKdV equation for surface waves as
\begin{equation}
2\eta_R^{(0)} + 3 \eta^{(0)}\eta_{\xi}^{(0)} + \frac{1}{3}\eta_{\xi\xi\xi}^{(0)} + \frac{\eta^{(0)}}{R} = 0. \label{Bouss_cKdV}
\end{equation}
Recalling that the depth-averaged Cartesian velocities are given by a linear combination of the polar velocity projections $U$ and $V$ by the formulae (\ref{cartv}), this gives the Cartesian velocity components to be
\begin{eqnarray}
u^{(0)} = \frac{x}{\sqrt{x^2 + y^2}} ~ \eta^{(0)}, \quad
v^{(0)} = \frac{y}{\sqrt{x^2 + y^2}} ~ \eta^{(0)}. \label{uv}
\end{eqnarray}

Finally, the KPII equation is derived using the change of variables $(t,x,y) \rightarrow (T,\xi,Y)$, where $T = \epsilon t$, $\xi = x - t$ and $Y = \sqrt{\epsilon} y$  {\color{black} (see, for example, \citealt{AS, J97} and references therein)}. We also scale the Cartesian velocity in the $y$-direction as $v \rightarrow \sqrt{\epsilon} v$ to remain consistent with the  coordinate scaling. Applying this change of variables to (\ref{Bouss1}) and (\ref{Bouss2}) gives
\begin{eqnarray}
&&u_{\xi} - \eta_{\xi} + \epsilon \left(v_Y + \eta u_{\xi} + u\eta_{\xi} + \eta_T \right) = O(\epsilon^2), \label{KPeqn1} \\
&&\eta_{\xi} - u_{\xi} + \epsilon \left(uu_{\xi} + \frac{1}{3}u_{\xi\xi\xi} + u_T \right) = O(\epsilon^2), \label{KPeqn2} \\
&&\eta_Y - v_{\xi} + \epsilon \left(uv_{\xi} + \frac{1}{3} u_{\xi\xi Y} + v_T \right) = O(\epsilon^2). \label{KPeqn3}
\end{eqnarray}
We now seek a solution of (\ref{KPeqn1})--(\ref{KPeqn3}) in the form:
\begin{equation}
\textcolor{black}{ \eta(T,\xi,Y) = \eta^{(0)}(T,\xi,Y) + \epsilon \eta^{(1)}(T,\xi,Y) + \textit{O}(\epsilon^2) \quad \mbox{as} \quad \epsilon \rightarrow 0, }
\end{equation} 
and similarly for the variables $u$ and $v$. We then consider solutions in the far field such that $T \sim \textit{O}(1)$, and  to leading order, $O(1)$, we obtain
\begin{eqnarray}
&&u^{(0)}_{\xi} = \eta^{(0)}_{\xi}, \label{KPlo1} \\
&&v^{(0)}_{\xi} = \eta^{(0)}_Y, \label{KPlo2}
\end{eqnarray}
such that $u^{(0)} = \eta^{(0)}$, assuming that any disturbances are generated only by the propagating waves. At the following order, $O(\epsilon)$, we obtain
\begin{eqnarray}
&&\eta_{\xi}^{(1)} - u_{\xi}^{(1)} = v_Y^{(0)} + \eta^{(0)}u_{\xi}^{(0)} + u^{(0)}\eta_{\xi}^{(0)} + \eta_T^{(0)}, \label{KPeps1} \\
&&u_{\xi}^{(1)} - \eta_{\xi}^{(1)} = u^{(0)}u_{\xi}^{(0)} + \frac{1}{3} u_{\xi\xi\xi}^{(0)} + u_T^{(0)}, \label{KPeps2}
\end{eqnarray}
from which, taking the sum of (\ref{KPeps1}) and (\ref{KPeps2}), and substituting (\ref{KPlo1}) gives 
\begin{equation}
2\eta^{(0)}_T + 3\eta^{(0)}\eta^{(0)}_{\xi} + \frac{1}{3}\eta^{(0)}_{\xi\xi\xi} + v^{(0)}_Y = 0. \label{nearlyKP}
\end{equation}
Taking the $\xi$-derivative of (\ref{nearlyKP}) and substituting the $Y$-derivative of (\ref{KPlo2}) gives the KPII equation for surface waves  
\begin{equation}
\left( 2\eta^{(0)}_T + 3\eta^{(0)}\eta^{(0)}_{\xi} + \frac{1}{3}\eta^{(0)}_{\xi\xi\xi}\right)_{\xi} + \eta^{(0)}_{YY} = 0. \label{KP}
\end{equation}
The slow time variable $T$ is more commonly used for the KPII equation and in the later sections makes comparisons to the solutions of the 2D Boussinesq--Peregrine equations simpler. {\color{black} For the KdV$\theta$ and cKdV equation the $R$-variable version could be more natural (see, for example, the discussion concerning the cKdV equation in \citealt{STCK}), but the slow time $T = \epsilon t$ version is easily recovered by the change of variables $(R, \xi, \theta) \to (T, \xi, \theta)$.}

% ---------- ---------- ---------- ---------- %
\section{{\color{black} The KdV$\theta$ equation and perturbed plane waves}}
\label{plane_waves}

{\color{black}In this section, we first derive the KdV$\theta$ equation  from the  2+1-dimensional amplitude equation (\ref{A})  for a stratified fluid with a parallel shear flow, which was previously derived  from the full set of Euler equations with the free-surface and flat-bottom boundary conditions  by \cite{KZ}. Since the general solution of the equation for the function $k$ is associated with plane waves \citep{HKG}, assuming that the waves exist,  it will be in the form  
\begin{equation}
k(\theta; a) = a \cos \theta + b(a) \sin \theta, 
\label{gen}
\end{equation}
with a problem-specific relation $b = b(a)$. Indeed, the waves associated with the general solution (\ref{gen}) have the fast variable 
\begin{equation}
\xi = r k(\theta) - s t = ax + b(a) y - st,
\end{equation}
with the constants $a$ and $b(a)$ having the meaning of the two wavenumbers of the plane wave propagating at an angle $\displaystyle \tan \varphi = b(a) / a $ to the flow (see figure \ref{angleflowpic}). The wavefronts are tangent to the ring wave associated with the corresponding singular solution, obtained as a geometrical envelope of the general solution (see figure \ref{sing_gen_sol_fig} for $u_0(z) = \gamma z, \ \gamma = {\rm const}$).  

\begin{figure}
	\centerline{\includegraphics[width=0.4\linewidth]{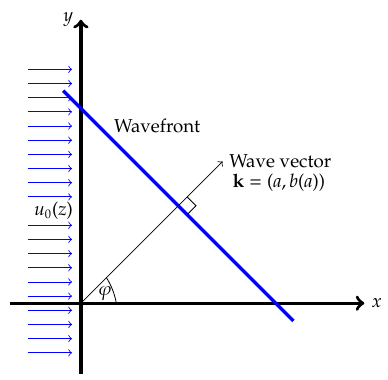}}
	\caption{Wavefront and wave vector of a plane wave propagating at an angle $\varphi$ to the direction of the shear flow.}
\label{angleflowpic}
\end{figure}
\begin{figure}
  \centerline{\includegraphics[width= 0.85 \linewidth]{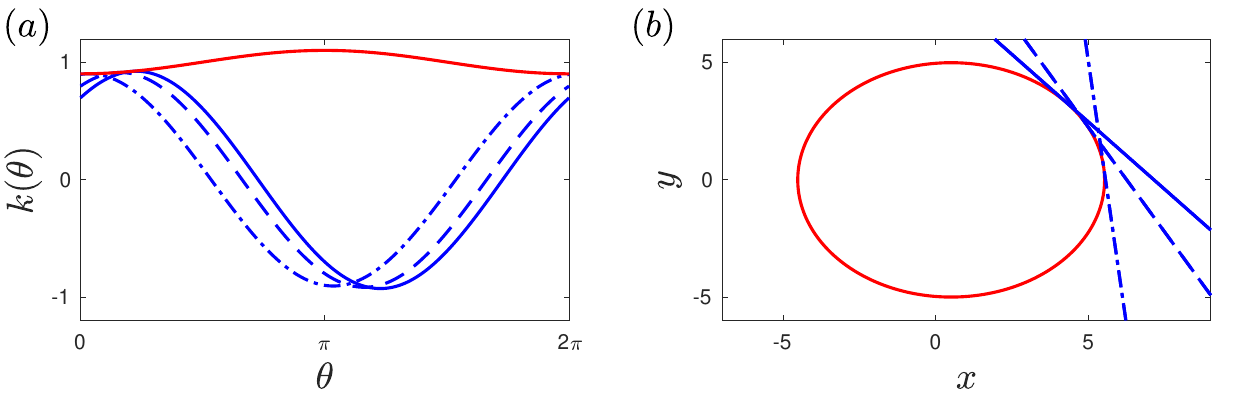}}
  \caption{$(a)$ The general solution (\ref{gs}) for $a = 0.7$ (blue), $a=0.8$ (dashed blue) and $a=0.9$ (dashed dot blue) and its envelope (the singular solution, solid red) all for $u_0(z) = \gamma z$, $\gamma = 0.2$. $(b)$ The wavefronts of the corresponding general and singular solutions from $(a)$ described by $rk(\theta) = 5$.}
	\label{sing_gen_sol_fig}
\end{figure}

Next, on the general solution (\ref{gen}), we obtain  $k^2 + k'^2 = a^2 + b(a)^2$ and $F = -s + [u_0(z) - c] a$. Hence, the modal  equations (\ref{modal1})--(\ref{modal3}) become $\theta$-independent, and therefore the modal function  is also $\theta$-independent, i.e. $\phi=\phi(z)$.  It then implies, using (\ref{c4}) and (\ref{3}), that $\mu_4 = 0$ for any stratification and any parallel shear flow $u_0(z)$. Hence, we obtain the amplitude equation, which has the form
\begin{equation}
\mu_1 A_R + \mu_2 A A_{\xi} + \mu_3 A_{\xi \xi \xi}  + \mu_5 \frac{A_\theta}{R} = 0, \label{A2}
\end{equation}
and the coefficients of this equation are computed using the general solution (\ref{gen}) in the formulae (\ref{c3}) and (\ref{c5}). Generally, the coefficient $\mu_5$ has a nontrivial dependence on $\theta$, while $\mu_1, \mu_2, \mu_3$ are some constants. For example, for surface waves in a homogeneous fluid over the linear shear flow $u_0(z) = \gamma z$ with $\gamma = {\rm const}$ we obtain
\begin{eqnarray}
\mu_1 &&= \frac{2s - \gamma a}{s (s-\gamma a)^2}, \
\mu_2 = \frac{[a^2+b(a)^2]^2(3s-3s\gamma a + \gamma^2 a^2)}{s^3(s-\gamma a)^3}, \
\mu_3 = \frac{[a^2+b(a)^2]^2}{3s^2}, \\
\mu_5 &&= \frac{[a \cos\theta + b(a)\sin\theta][2b(a)(s-\gamma a) \cos\theta - (2as + a^2 \gamma - b(a)^2 \gamma)\sin \theta]}{s(s-\gamma a)^2}.
\end{eqnarray}
When there is no shear flow the coefficients reduce to $\mu_1 = 2$, $\mu_2 = 3$, $\mu_3 = 1/3$, and $\mu_5 = -2\sin\theta\cos\theta$, resulting in the equation
\begin{equation}
A_R + \frac 32 A A_{\xi} + \frac 1 6 A_{\xi \xi \xi}  - \frac 12 \sin  2 \theta  \frac{A_\theta}{R} = 0, \label{KdVTHETA}
\end{equation}
which coincides with equation (\ref{KdVT}) derived in Section 2 directly from the 2D Boussinesq--Peregrine system. If the wave has a plane wavefront, then $A_{\theta} = 0$ (there is no change in the tangential direction along the wavefront), and the amplitude equation (\ref{A2}) (and hence (\ref{KdVTHETA})) has a reduction to the KdV equation (see Appendix \ref{appC}). However, the full equation (\ref{A2}), and its particular case (\ref{KdVTHETA}), are more general. To the best of our knowledge, equations (\ref{A2}) and  (\ref{KdVTHETA}) have not been derived before and we will refer to both as the KdV$\theta$ equation.}

In this section we {\color{black} consider plane waves subject to long transverse perturbations. As an example, we  apply  long transverse amplitude perturbations of finite strength} to the line-soliton solution of the KdV$\theta$ equation (\ref{KdVTHETA}), where we have changed the variables from $(R,\xi,\theta)$ to $(T=\epsilon t,\xi,\theta)$ for {\color{black} a more accurate comparison of numerical solutions of the reduced amplitude equation} to those of the 2D Boussinesq--Peregrine system. {\color{black} To leading order, the change of variable $R \rightarrow T$ does not change the form of the amplitude equation (\ref{A2}), provided we map $\mu_1 \to \mu_1 / s$ and $\mu_5 \to \mu_5 / s$:
\begin{equation}
\mu_1  A_T +  s \mu_2 A A_{\xi} + s \mu_3 A_{\xi \xi \xi} + \mu_5 \frac{A_{\theta}}{T} = 0.
\label{A2T}
\end{equation}
The KdV$\theta$ equation (\ref{A2T})  can be analysed analytically, using a mapping similar to that noted by \citet{J90} for ring waves.  Indeed, for every value of $\theta$, the change of variables
\begin{equation}
\tilde{\theta} = T \exp \left( - \int \frac{\mu_1}{\mu_5} ~ \mathrm{d}\theta \right)
\label{4}
\end{equation}
maps  the KdV$\theta$ equation (\ref{KdV})  to the usual KdV equation
\begin{equation}
A_T + \frac{s \mu_2}{\mu_1}  AA_{\xi} +  \frac{s \mu_3}{\mu_1}  A_{\xi\xi\xi} = 0, \label{kdveqn1}
\end{equation}
since equations (\ref{c3})  and (\ref{c5}) imply that  the only $\theta$-dependent coefficient in (\ref{A2T}), and the slow $R$ version (\ref{A2}), is $\mu_5$.
However, the initial condition for the KdV equation (\ref{kdveqn1}) will have parametric dependence on $\tilde\theta$, and hence, on the polar angle $\theta$. 

Here, we consider surface waves without a shear flow. Then $s = 1$, and the KdV$\theta$ equation (\ref{A2T}),} for the waves propagating in the $x$-direction, takes the form
\begin{eqnarray}
&&A_T + \frac{3}{2}  AA_{\xi} + \frac{1}{6}  A_{\xi\xi\xi} -  \frac 12 \sin 2 \theta   \frac{A_{\theta}}{T} = 0, \label{KdV} 
\end{eqnarray}
where
\begin{eqnarray}
&&T = \epsilon t, \quad \xi = x-t, \quad \theta = \arctan \frac{y}{x}.
\end{eqnarray}
A long transverse amplitude perturbation is added to the initial condition, at some $T = T_0$, by multiplying the initial condition by a suitable function of $\theta$. The unperturbed initial condition  is the $\theta$-independent exact line-soliton solution of the KdV$\theta$ equation (\ref{KdV}), given by 
\begin{equation}
{\color{black} A_0(T_0,\xi,\theta) = 2 \tilde v \operatorname{sech}^2 \left(\frac{1}{2} \sqrt{6 \tilde v} \left[\xi - \tilde v T_0 - \xi_0 \right] \right),}
\end{equation}
{\color{black} where $\tilde v > 0$ and $\xi_0$ are some constants.
We take the perturbed initial condition to be in the form
\begin{equation}
A(T_0,\xi,\theta) = \left(1 + \alpha \operatorname{sech}^2 \beta\theta \right) A_0(T_0,\xi,\theta), \label{pert_soliton} 
\end{equation}
although we could choose to multiply by any other function of $\theta$.} Here, the constants $\alpha$ and $\beta$ control the amplitude and width of the perturbation, $\tilde v$ is the speed of the soliton, and $\xi_0$ is an initial phase shift. The multiplication of the original $\operatorname{sech}^2$ soliton by this $\theta$-dependent function generates a smooth, localised perturbation in the central region of the wavefront. 

{\color{black} The mapping (\ref{4}) yields
\begin{equation}
\tilde \theta = T \tan \theta,
\end{equation}
and the KdV equation (\ref{kdveqn1}) simplifies to become
\begin{equation}
A_T + \frac 32 AA_{\xi} +  \frac 16 A_{\xi\xi\xi} = 0, \label{kdveqn}
\end{equation}
which we now need to solve for the initial condition
\begin{equation}
A(T_0,\xi, \tilde \theta) = \left[1 + \alpha \operatorname{sech}^2 \left (\beta \arctan \frac{\tilde \theta}{T_0}\right ) \right ] A_0 \left( T_0,\xi, \arctan \frac{\tilde \theta}{T_0} \right).
\end{equation}}
Next, to use the results of the Inverse Scattering Transform \citep{GGKM}, we cast the KdV equation (\ref{kdveqn}) into the canonical form
\begin{equation}
U_{\tau} - 6UU_X + U_{XXX} = 0, \label{cankdv}
\end{equation}
via the scalings $\displaystyle U = - \frac 32 A,  \ \tau =  \frac 16 T,  \ X = \xi.$ The wavefield is then defined by the spectrum of the Schr\"{o}dinger equation
\begin{equation}
\Psi_{XX} + \left[\lambda - U(\tau_0,X;\tilde\theta) \right]\Psi = 0, \label{Schrodinger_eq}
\end{equation}
where the potential is given by the perturbed initial condition (\ref{pert_soliton}) 
\begin{eqnarray}
&&U(\tau_0,X;\tilde\theta) = - \Lambda \operatorname{sech}^2 \left(\frac{{\color{black} X - 6 \tilde v \tau_0 - \xi_0}}{L} \right),
\end{eqnarray}
where
\begin{eqnarray}
&&\Lambda = 3 \tilde v \Delta, \ \ L = \frac{2}{\sqrt{6 \tilde v}},  \ \ \Delta\left(\tau_0;\tilde\theta \right) = 1 + \alpha \operatorname{sech}^2\left( \beta \arctan \frac{\tilde{\theta}}{6\tau_0} \right),
\end{eqnarray}
{\color{black} i.e. we have different initial conditions for different values of $\tilde \theta$, and, hence, for different values of the polar angle $\theta$.}
The number of generated solitons depends on $\theta$ and  is given by the greatest integer satisfying the inequality
\begin{equation}
N\left(\tau_0;\tilde\theta\right) < \frac{1}{2} \left[\left(1 + 4\Lambda L^2 \right)^{\frac{1}{2}} + 1 \right]
\end{equation}
(see, e.g. \citet{LL}). The discrete eigenvalues are given by
\begin{eqnarray}
&&\lambda = -k_n^2,
\end{eqnarray}
where
\begin{eqnarray}
&&k_n\left(\tau_0;\tilde{\theta}\right) = \frac{1}{2L}\left[\left(1 + 4\Lambda L^2\right)^{\frac{1}{2}} - (2n-1) \right] > 0, \quad n = 1,2,\ldots,N. \label{ISTk}
\end{eqnarray}
For positive $\alpha$- and $\tilde v$-values, the central perturbed region fissions into at least two solitons and a dispersive wave train. However, for small $\alpha$-values the secondary soliton is small and for a very long time  it blends with the radiation. The outer, unperturbed section of the line soliton propagates as a single {\color{black} curved} soliton. We therefore expect a localised 2D region of radiation behind a smaller soliton in the wake of the main perturbed soliton. The long-time asymptotics takes the form
 \begin{equation}
U\left(\tau,X;\tilde{\theta}\right) \simeq - \sum_{n=1}^N 2k_n^2 \operatorname{sech}^2 \left( k_n(X - 4k_n^2\tau - X_n)\right) + \text{radiation}, \label{ISTU}
\end{equation}
(e.g. \citealt{D}), where the phase shifts are given by
\begin{eqnarray}
X_1 = \frac{1}{2k_1} \ln \left(\frac{c_1}{2k_1} \right), \quad
X_n = \frac{1}{2k_n} \ln \left(\frac{c_n}{2k_n} \prod_{m=1}^{n-1} \left( \frac{k_n - k_m}{k_n + k_m} \right)^2 \right), \quad n > 1,
\end{eqnarray}
and the constants $c_n$ are found as
\begin{equation}
c_n = \left(\int_{-\infty}^{\infty} \Psi_n^2(x) \text{ d}x \right)^{-1},
\end{equation}
where $\Psi_n(X)$ is the eigenfunction of (\ref{Schrodinger_eq}) corresponding to the $n^{\text{th}}$ {\color{black} eigenvalue,  $-k_n^2$:}
\begin{equation}
\Psi_n = \text{const}\left(1 - \tanh^2 \frac{x}{L} \right)^{\frac{k_nL}{2}} {}_{2}F_{1}\left(1-n, 2k_nL + n,k_nL+1, \frac{1 - \tanh \frac{x}{L}}{2} \right).
\end{equation}
Here, ${}_{2}F_{1}(\cdots)$ is the hypergeometric function, and the constant should be chosen to normalise the eigenfunction at  infinity,
\begin{equation}
\Psi_n \sim \exp^{-k_nx} \quad \text{as} \quad x \rightarrow + \infty.
\end{equation}
We can therefore analytically describe the main soliton and the extra solitons produced, for every direction defined by the polar angle $\theta$.

\begin{figure}
  \centerline{\includegraphics[width=  \linewidth]{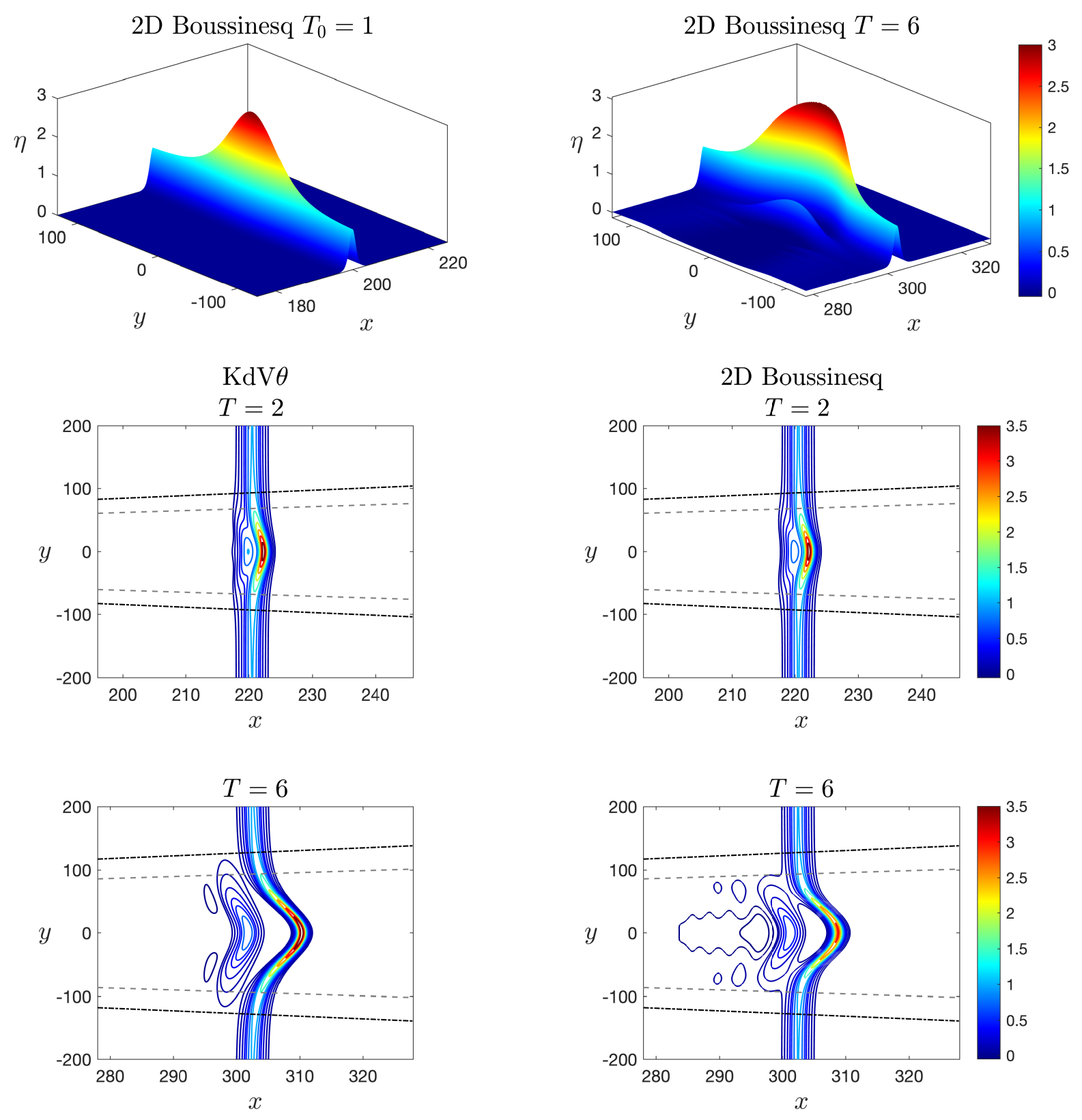}}
  \caption{Plots of the numerical solution of both the KdV$\theta$ equation and 2D Boussinesq--Peregrine system for a perturbed line soliton with the parameters $\epsilon = 0.05$, $\tilde v = 0.5$, $\xi_0 = 180$, $\alpha = 2$, $\beta = 5$ between $T_0 = 1$ and $T = 6$. The dark grey (dash-dotted) and the light grey (dashed)  lines illustrate theoretical (IST) predictions for the perturbed regions of the main and secondary solitons,  respectively.}
	\label{pert_comps} 
\end{figure}

When there is no perturbation, i.e. $\Delta = 1$, the solution is simply the exact line-soliton solution. When $\Delta > 1$ there is a localised perturbation to the wavefront, detectable in some region given by $-\tilde{\theta}_1 \lesssim \tilde{\theta} \lesssim \tilde{\theta}_1$ such that the amplitude $A(T,\xi,\theta)  \gtrsim  A(T_0,\xi,\theta) + \epsilon$. The estimate for the region determining where the perturbation lies is therefore given by the formula $4 k_1(\tilde\theta)^2/3 \gtrsim - 2 U(\tau_0,X;\tilde\theta)/3 + \epsilon$ arising from (\ref{ISTU}). The small secondary solitons are then detectable when $A(T,\xi,\theta) \gtrsim \epsilon$ and hence when $4 k_n(\tilde\theta)^2/3 \gtrsim \epsilon$. {\color{black} These regions are shown  in figure \ref{pert_comps}, where we compare the performance of the KdV$\theta$ equationl with the results of direct numerical simulations for the parent system.}

The corresponding initial condition in the $xy$-plane for the 2D Boussinesq--Peregrine system (\ref{Bouss1})--(\ref{Bouss2}) is obtained using the initial condition to the KdV$\theta$ equation (\ref{pert_soliton}):
\begin{eqnarray}
&&\eta(t_0,x,y) = 2 \tilde v \left(1 + \alpha \operatorname{sech}^2 \left( \beta \arctan \frac{y}{x} \right) \right) 
 \operatorname{sech}^2 \left(\frac{1}{2} \sqrt{6 \tilde v} \left(x - (1 + \epsilon  \tilde v)t_0 -  \xi_0 \right) \right), \qquad \label{Bouss_pert_soliton1} \\
&&u(t_0,x,y) = \eta(t_0,x,y), \quad
v(t_0,x,y) = 0. \label{Bouss_pert_soliton3}
\end{eqnarray}
where $\xi_0 = {\rm const}$.
The computation parameters can be found in table~\ref{table_soliton_perts}, and the corresponding numerical results are shown in figure \ref{pert_comps}. 

\begin{table}
  \begin{center}
\def~{\hphantom{0}}
  \begin{tabular}{cccccc}
      Regime      & $x \times y$			        & $N_x$     & $N_y$        & $\Delta_t$ \\ [5pt]
      Boussinesq  & $[150,400] \times [-250,250]$   & 2500      & 4000         & $10^{-2}$  \\ [10pt]
      Regime      & $\xi \times \theta$			    & $N_{\xi}$ & $N_{\theta}$ & $\Delta_T$ \\ [5pt]
	  KdV$\theta$ & $[0,200] \times [-\pi/4,\pi/4]$ & 2000      & 500          & $5 \times 10^{-4}$
  \end{tabular}
  \caption{Computation parameters for the KdV$\theta$ equation and 2D Boussinesq--Peregrine system to generate figure \ref{pert_comps}.}
  \label{table_soliton_perts}
  \end{center}
\end{table}

{\color{black} We solve the 2D Boussinesq--Peregrine system for $t \in [20,120]$ and $\epsilon = 0.05$, where comparison is made to the KdV$\theta$ equation that is solved for $T \in [1,6]$. Comparisons can therefore be made at the same time.} There is good quantitative agreement between the two solutions around the slow time value  $T = 2$. The second soliton is emerging, and there is also some extra radiation in the 2D Boussinesq--Peregrine system. In longer runs, shown for the slow time value $T = 6$, the agreement between the KdV$\theta$ equation and the 2D Boussinesq--Peregrine equations worsens. The main perturbed soliton starts to split sideways, spreading out across the wavefront and in doing so reduces in amplitude in the centre. This feature is not seen in the KdV$\theta$ equation. The second soliton has now fissioned from the original soliton, {\color{black} and the analytical estimates based on the IST give good predictions for the perturbed regions of the main and secondary solitons.} In the 2D Boussinesq--Peregrine equations, the second soliton has radiated more than in the KdV$\theta$ regime, leading to the noticeable difference. However, there is still good qualitative agreement between the full parent system and the reduced model, despite the strong perturbation and long propagation time.

\begin{figure}
  \centerline{\includegraphics[width= 0.8\linewidth]{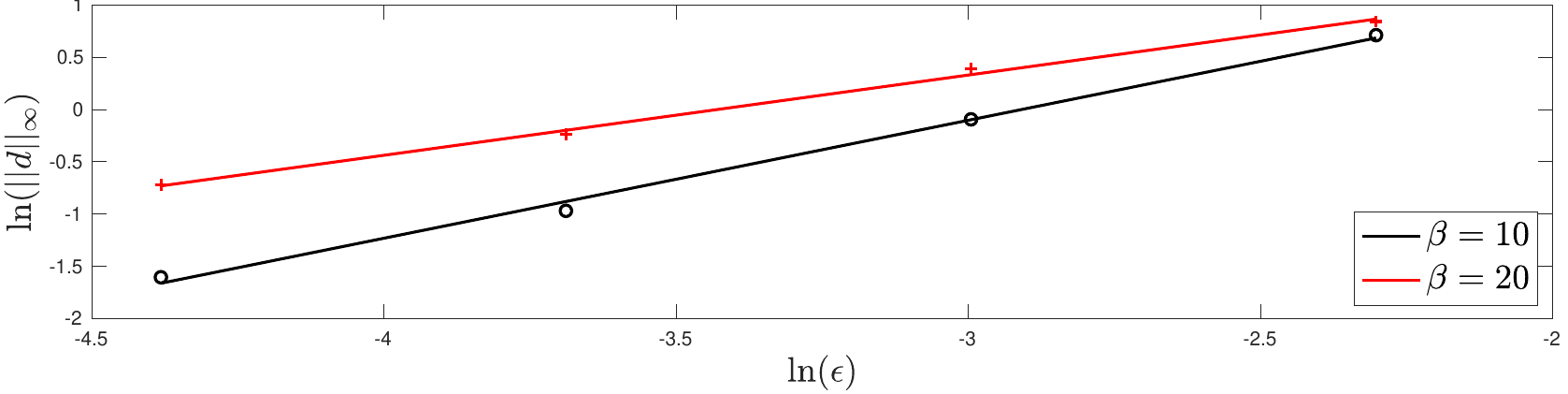}}
  \caption{Convergence of {\color{black} the solution of  the KdV$\theta$ equation to the solution  of the } 2D Boussinesq--Peregrine system for different lengths of perturbation and $\epsilon$, where {\color{black}  $||d||_\infty$ is the $L^\infty$ norm of the difference between solutions of the 2D Boussinesq--Peregrine system and KdV$\theta$ equation,} and $T \in [1,2]$, $\tilde v = 0.5$, $\xi_0 = 200$, and $\alpha = 0.5$.}
	\label{Pert_Convergence}
\end{figure}

Eventually, all strong, long transverse amplitude perturbations split and travel sideways along the wavefront  with two ring wave arcs being radiated as the perturbation travels bidirectionally along the wavefront. For long enough perturbations, {\color{black}  the $L^{\infty}$ norm  $||d||_\infty$  of the difference between solutions of} the 2D Boussinesq--Peregrine system and  the KdV$\theta$ equation is consistent with the model and scales as $\textit{O}(\epsilon)$. The computation parameters in table~\ref{table_soliton_convergence} correspond to figure \ref{Pert_Convergence}. In figure 5, for the strong perturbation of $\alpha = 0.5$ considered here and the perturbation width parameter $\beta = 20$ the fitted line has the gradient $0.76$. When the perturbation is wider, $\beta = 10$, the fitted line has a greater gradient 1.1. Therefore, for sufficiently long perturbations the KdV$\theta$ equation is accurate to $\textit{O}(\epsilon)$. We note that the KPII equation is valid too in this regime, as well as for narrower perturbations. However, the KdV$\theta$ equation is a much simpler model, which can be used {\color{black} to describe the intermediate asymptotics of plane waves subject to sufficiently long transverse perturbations of finite amplitude.}

\begin{table}
  \begin{center}
\def~{\hphantom{0}}
  \begin{tabular}{ccccc}
      Regime      & $x \times y$			        & $N_x$     & $N_y$        & $\Delta_t$ \\[5pt]
      Boussinesq  & $[140,280] \times [-125,125]$   & 1400      & 2500         & $10^{-2}$ \\ [10pt]
      Regime      & $\xi \times \theta$			    & $N_{\xi}$ & $N_{\theta}$ & $\Delta_T$ \\[5pt]
	  KdV$\theta$ & $[100,300] \times [-\pi/4,\pi/4]$ & 2000      & 500        & $5 \times 10^{-4}$
  \end{tabular}
  \caption{Computation parameters for the KdV$\theta$ equation and 2D Boussinesq--Peregrine {\color{black} system} to generate figure \ref{Pert_Convergence}.}
  \label{table_soliton_convergence}
  \end{center}
\end{table}

% ---------- ---------- ---------- ---------- %
\section{{\color{black} Perturbed ring waves}}
\label{ring_waves}

{\color{black} In this section, we extend the studies of concentric ring waves by \citet{CW, STCK} to non-axisymmetric perturbed ring waves, within the framework of the 2D Boussinesq--Peregrine system (\ref{Bouss1})--(\ref{Bouss2}). We use the cKdV equation (\ref{Bouss_cKdV}) and the related formulae (\ref{uv})  for velocity components to define the initial conditions for our numerical runs, and consider both localised and periodic perturbations of concentric waves.}

{\color{black} We consider a ring wave which is placed sufficiently far away from the origin. Then, the cylindrical divergence/convergence term  in the cKdV equation (\ref{Bouss_cKdV}) can be treated as a perturbation of the KdV equation, and one can consider the axisymmetric solution initiated by the KdV soliton bent into a ring wave (see \citealt{KoK, DPS, J99, STCK}):}
\begin{equation}
\eta(t_0,x,y) = 2 \tilde v \operatorname{sech}^2 \left( \frac{1}{2} \sqrt{6 \tilde v} \left[\sqrt{x^2 + y^2} - (1 + \epsilon \tilde v)t_0 - {\color{black} r_0} \right] \right).
\end{equation}
{\color{black} Here, $\tilde v > 0$ and $r_0$ are some constants. The Cartesian velocities $u$ and $v$ are defined using the related approximations given by the formulae (\ref{uv}). 

{\color{black} Next,  we consider the initial condition with a localised perturbation for the Boussinesq--Peregrine system (\ref{Bouss1})--(\ref{Bouss2})  in the form}}
\begin{eqnarray}
&&\eta(t_0,x,y) = 2 \tilde v \left(1 + \alpha H(x) \operatorname{sech}^2 \beta y \right) \operatorname{sech}^2\left( \frac{1}{2} \sqrt{6 \tilde v} \left[\sqrt{x^2 + y^2} - (1 + \epsilon \tilde v)t_0 - {\color{black} r_0} \right] \right), \qquad \\
&&u(t_0,x,y) = \frac{x}{\sqrt{x^2 + y^2}} ~ \eta(t_0,x,y), \quad
v(t_0,x,y) = \frac{y}{\sqrt{x^2 + y^2}} ~ \eta(t_0.x,y), \label{vel}
\end{eqnarray}
where $\alpha$ and $\beta$ control the amplitude and width of the perturbation, respectively, and $H = H(x)$ is the Heaviside function to only give the perturbation on one side of the ring. {\color{black} 
At the origin of the $xy$-plane, we set $\eta(t_0,0,0) = u(t_0,0,0) = v(t_0,0,0) = 0$ to avoid division by zero, 
and, for all ring-wave simulations, we use the computation parameters in table~\ref{table_parameters2}. }

\begin{table}
  \begin{center}
\def~{\hphantom{0}}
  \begin{tabular}{cccccc}
      Direction & $x \times y$			 		 & $N_x$ & $N_y$ & $\Delta_t$ \\[5pt]
      Outward   & $[-200,200] \times [-200,200]$ & 2800  & 2800  & $10^{-2}$ \\
      Inward    & $[-200,200] \times [-200,200]$ & 2800  & 2800  & $10^{-2}$
  \end{tabular}
  \caption{Computation parameters for the 2D Boussinesq--Peregrine {\color{black} system} simulations of ring waves.}
  \label{table_parameters2}
  \end{center}
\end{table}

{\color{black} For the outward-propagating ring wave we solve the 2D Boussinesq--Peregrine system  for $t \in [0,150]$, $\epsilon = 0.01$, $r_0 = 20$, $\tilde v = 0.5$, and $\alpha = 0.5$ for the two values of $\beta$ shown, and the computation parameters in table~\ref{table_parameters2} give figure \ref{Outward_ring_b=2,20}.}

\begin{figure}
  \centerline{\includegraphics[width=  0.9 \linewidth]{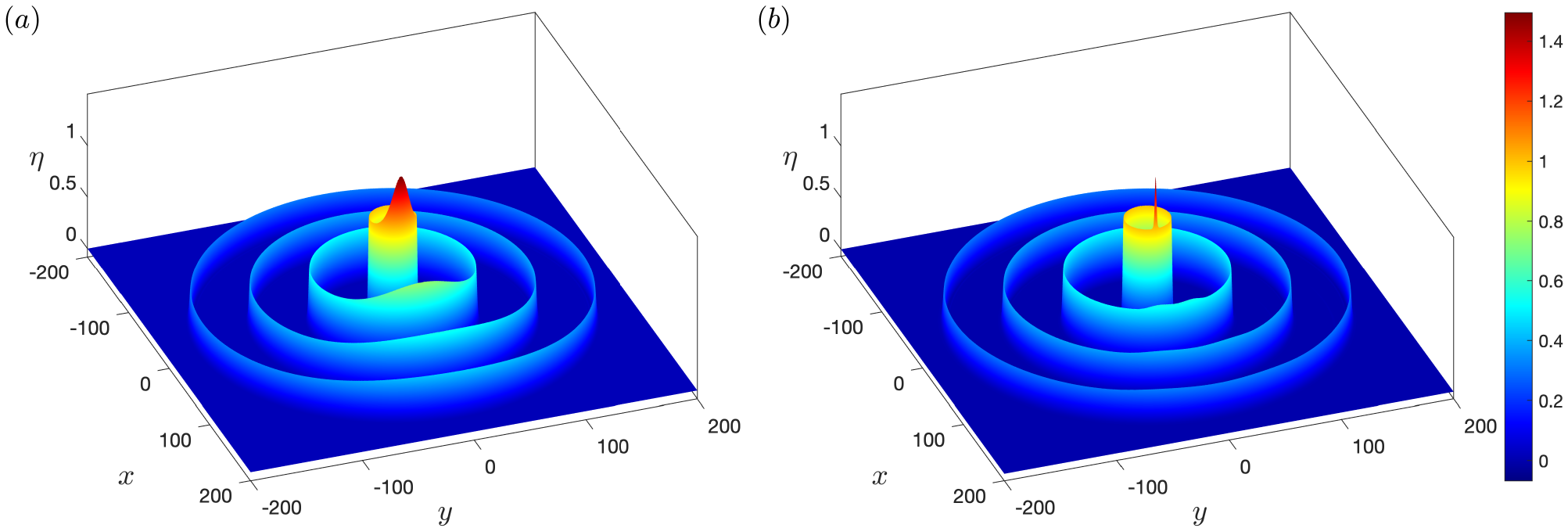}}
  \caption{3D plots of an outward-propagating, perturbed ring wave plotted at the times $t = 0,50,100,150$, where (\textit{a}) $\beta = 2$ and (\textit{b}) for $\beta = 20$.}
	\label{Outward_ring_b=2,20} 
\end{figure}

{\color{black} We consider two perturbations. The wider perturbation, with $\beta = 2$, has a width  comparable to the radius of the initial ring. It expands and decreases in amplitude as the wave radially diverges. The narrower perturbation, with $\beta = 20$, has a width comparable to the wavelength of the concentric wave. It instead rapidly splits and travels sideways along the wavefront, eventually becoming unnoticeable. Both outward-propagating perturbed ring waves are stable.}

\begin{figure}
  \centerline{\includegraphics[width= 0.9  \linewidth]{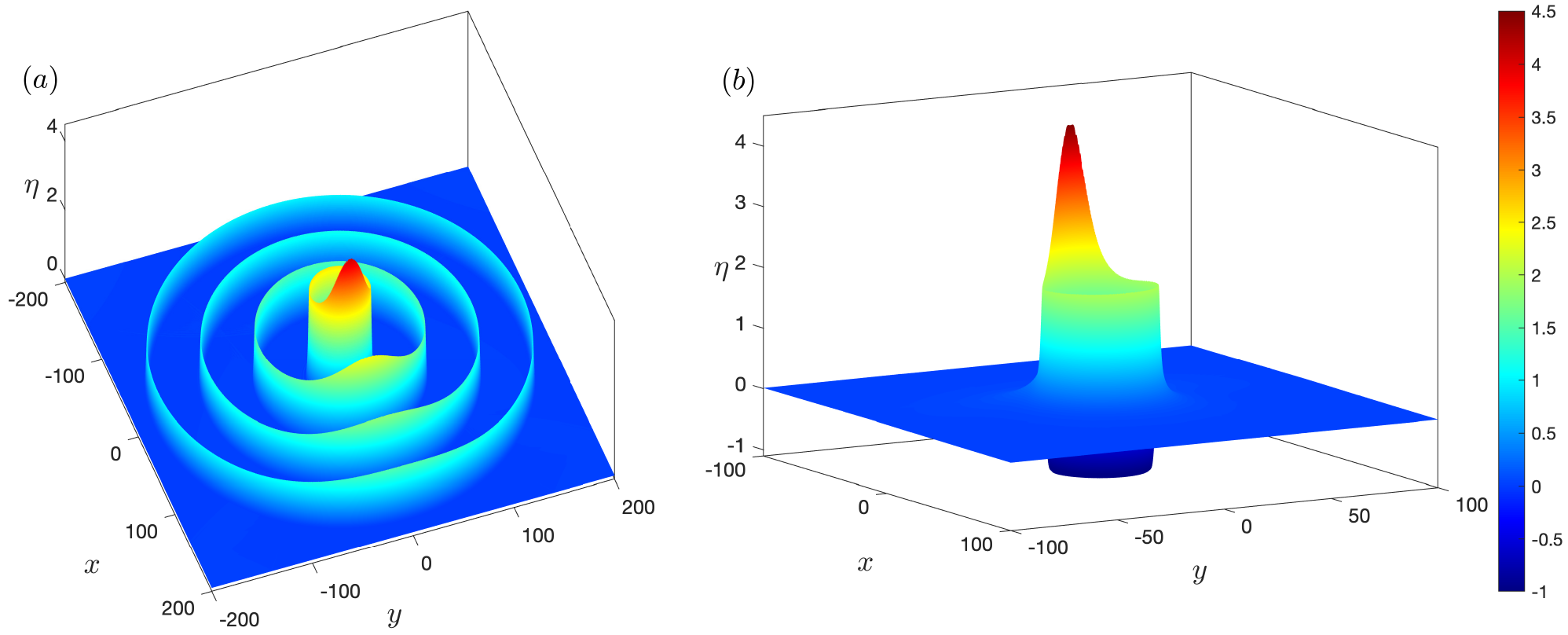}}
  \caption{2D plot of an inward-propagating perturbed ring wave for the perturbation parameters $\alpha = 0.5$ and $\beta = 2$, using $\epsilon = 0.01$ and plotted at the times (\textit{a}) $t=0,50,100,150$, and (\textit{b}) $t = 200$.}
	\label{Inward_ring_b=2} 
\end{figure}

\begin{figure}
  \centerline{\includegraphics[width= 0.9  \linewidth]{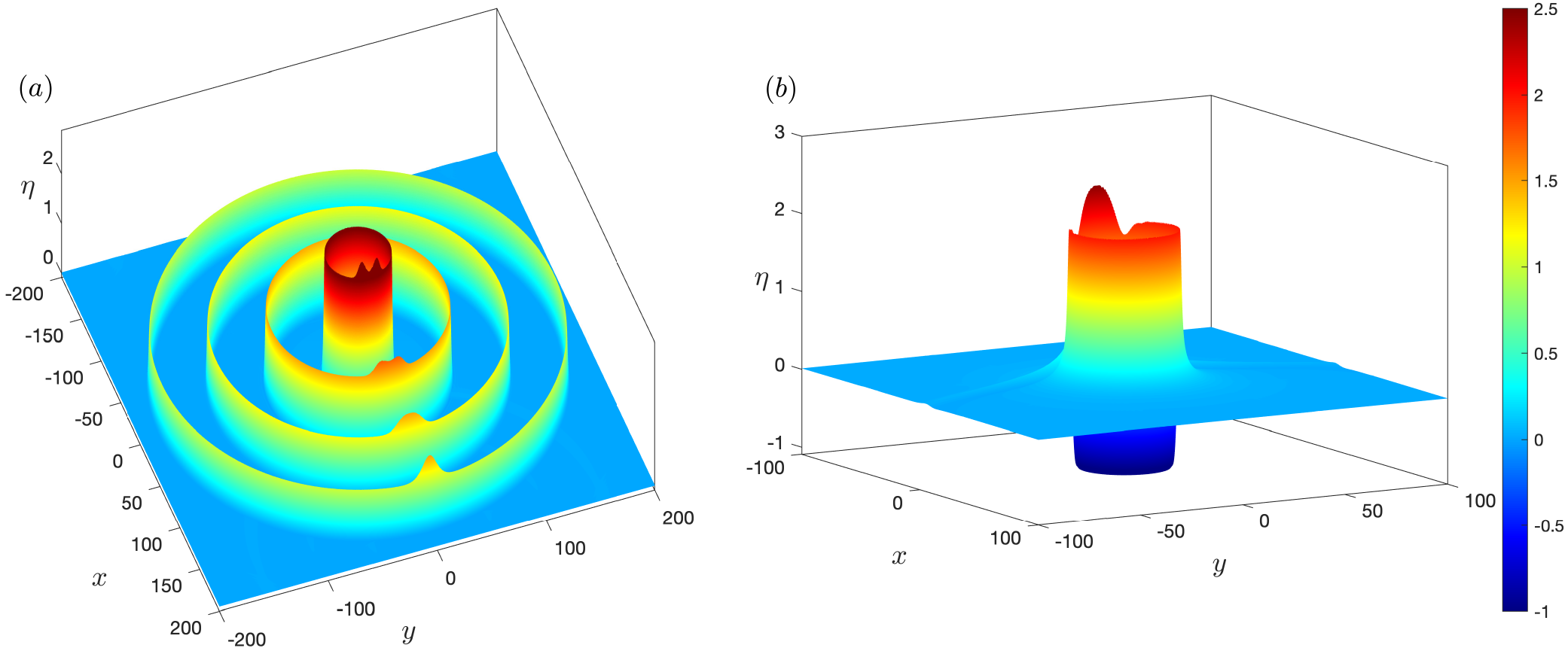}}
  \caption{2D plot of an inward-propagating perturbed ring wave for the perturbation parameters $\alpha = 0.5$ and $\beta = 20$, using $\epsilon = 0.01$ and plotted at the times (\textit{a}) $t=0,50,100,150$, and (\textit{b}) $t = 200$.}
	\label{Inward_ring_b=20} 
\end{figure}

To generate inward-propagating ring waves in the 2D Boussinesq--Peregrine system, we multiply the velocities (\ref{vel}) by $-1$, to reverse the direction of propagation. {\color{black} We then solve this system for $t \in [0,250]$, $\epsilon = 0.01$,  $r_0 = 180$, $\tilde v = 0.5$, and $\alpha = 0.5$ for the two values of $\beta$ shown, and the computation parameters listed in table~\ref{table_parameters2}. The results are shown in figures \ref{Inward_ring_b=2} and \ref{Inward_ring_b=20}. }

\begin{figure}
  \centerline{\includegraphics[width=  0.9 \linewidth]{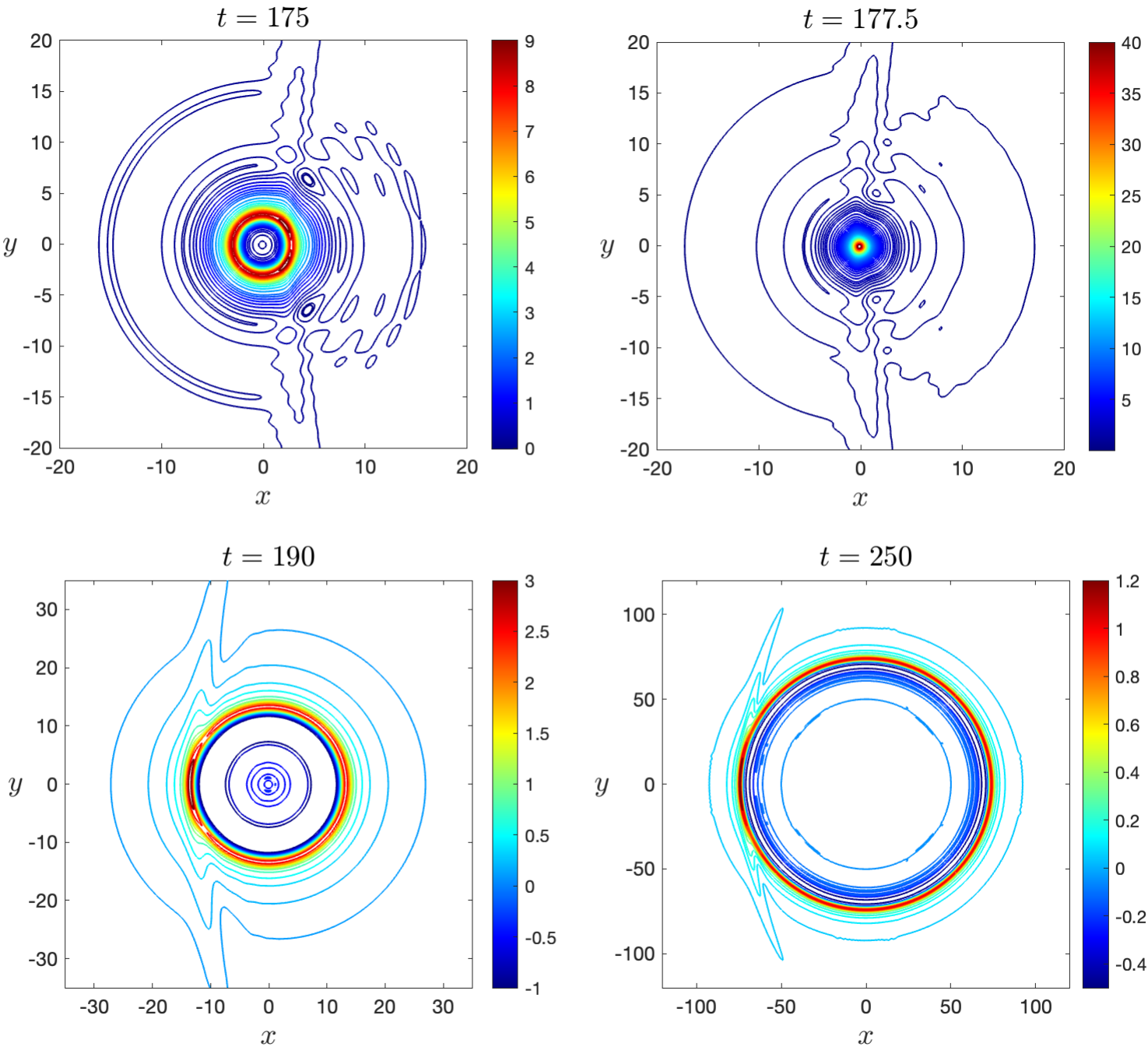}}
  \caption{Contour plots of the inward-propagating perturbed ring wave from figure \ref{Inward_ring_b=20} at before, during, and after the propagation through the origin. The perturbation parameters $\alpha = 0.5$ and $\beta = 20$, using $\epsilon = 0.01$ and plotted at the times $t = 175, 177.5, 190, 250$.}
	\label{Inward_ring_contour} 
\end{figure}

The wider perturbation, with $\beta = 2$,  comparable to the radius of the ring,  focuses to a point without splitting. The narrower perturbation, with $\beta = 20$, comparable to the wavelength, splits, taking longer to do so than its outward-propagating counterpart. In both figures \ref{Inward_ring_b=2} and \ref{Inward_ring_b=20}, the wave propagates through the origin and reforms the general shape as before, with the perturbation seemingly remaining on the same path, now as an outward-propagating perturbed ring wave. The case of the narrower  ($\beta = 20$) perturbation has more radiation on the trailing edge of the wave compared with the wider ($\beta = 2$) perturbation case. {\color{black} When the perturbation is narrower, and once it has passed through the origin, the perturbation starts to deform the shape of the ring, as seen in figure \ref{Inward_ring_contour}.  However, when the wave is at the origin, with the maximum amplitude $||\eta ||_{\infty} = 44.12$ occurring at $t = 177.5$, the wave is almost concentric. Once the wave has passed through the origin, it retains the form of the perturbation, now as an outward-propagating perturbed ring wave. The wavefront is constantly radiating small waves, and the moustache-shaped radiation from the perturbation is clearly visible in both figures \ref{Inward_ring_b=20} and \ref{Inward_ring_contour}. Once the wave has propagated sufficiently far outward, it retains its concentric shape.}

\begin{figure}
  \centerline{\includegraphics[width= 0.9 \linewidth]{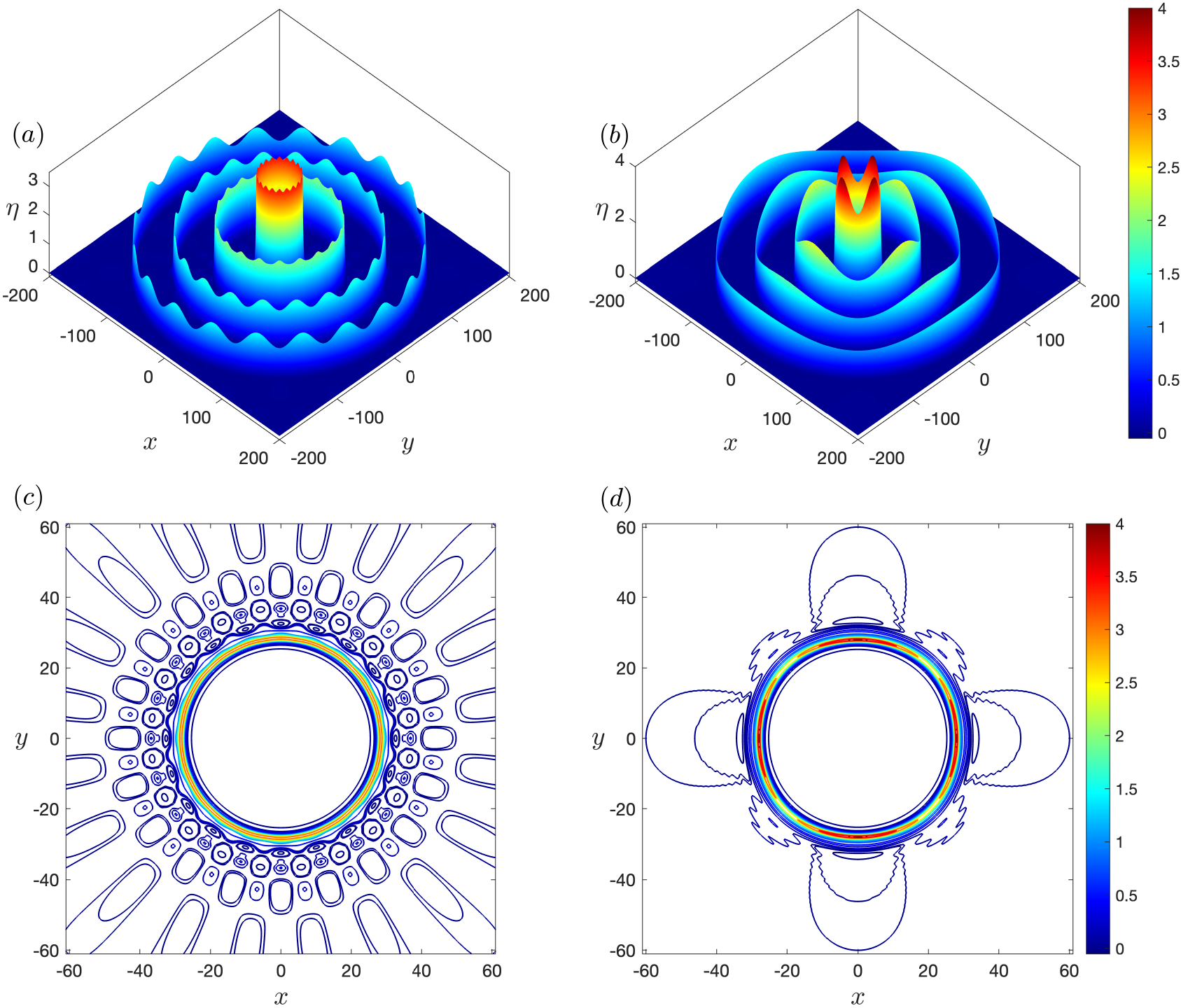}}
  \caption{Numerical solution of an inward-propagating periodically perturbed ring wave where $\epsilon = 0.01$, $r_0 = 180$, $\tilde v = 0.5$ and $\alpha = 0.25$ where the 3D plots (\textit{a}) $\beta = 20$ and (\textit{b}) $\beta = 4$ are plotted at $t = 0, 50, 100, 150$. The corresponding contour plots (\textit{c}) and (\textit{d}) are plotted at $t = 150$.}
	\label{Inward_ring_periodic} 
\end{figure}

We now apply a periodic perturbation to the inward-propagating ring wave. {\color{black} Theoretical considerations by \citet{OS, SP, P91} suggest that there exists a critical wavelength of the perturbation, such that  above this wavelength, the perturbed ring wave is unstable, whereas below  this critical value, it is stable. However, an analytical formula for this critical wavelength is not available at present, and deriving it is a separate problem beyond the scope of the present study. Here, in the first numerical study of the problem, we aim to provide an illustrative example of this qualitative difference.} Taking the initial condition
\begin{eqnarray}
&& \eta(t_0,x,y) = 2 \tilde v \left(1.25 +  \alpha \cos \left( \beta \arctan \frac{y}{x} \right)\right) \operatorname{sech}^2\left( \frac{1}{2} \sqrt{6 \tilde v} \left[\sqrt{x^2 + y^2} - x_0 \right] \right),  \\
&&u(t_0,x,y) = -\frac{x}{\sqrt{x^2 + y^2}} ~ \eta(t_0,x,y), \quad
v(t_0,x,y) = -\frac{y}{\sqrt{x^2 + y^2}} ~ \eta(t_0,x,y),
\end{eqnarray}
we solve the 2D Boussinesq--Peregrine system {\color{black} (\ref{Bouss1})--(\ref{Bouss2}) for $t \in [0,150]$, $\epsilon = 0.01$, $\tilde v = 0.5$, and $\alpha = 0.5$ for $\beta$ and $\beta = 20$ and $\beta=4$, and the computation parameters listed in table~\ref{table_parameters2}}. The results are shown in figure \ref{Inward_ring_periodic}.

{\color{black} It is clear that the higher-frequency periodic perturbation does not grow  faster than the rate of the radial convergence}, whereas the low-frequency counterpart does. Both cases produce periodic radiation patterns around the ring. However, the higher-frequency perturbation radiates considerably more, as seen in figure \ref{Inward_ring_periodic}, but remains largely concentric, in contrast to the low-frequency perturbation, which notably deforms the ring. {\color{black}  Hence, our simulations show significant qualitative differences in the behaviour of these two perturbations.}

% ---------- ---------- ---------- ---------- %
\section{{\color{black} Pure and perturbed hybrid waves}}
\label{hybrid_waves}

In this section, we construct and model hybrid waves, {\color{black} generated by initial conditions formed by an arc of a ring wave and two tangent plane waves  \citep{OS,HKG}. The two regions are constructed using the same description for plane and ring waves as above, and the wavefront has a continuous derivative.} Surface signatures of similarly looking internal waves generated in the Strait of Gibraltar can be seen {\color{black} in figure 1, as well as} figures 13 and 14 of \citet{A}. As discussed in the Introduction, all considerations of our present work can be extended to internal waves {\color{black} and to parallel shear flows.}

{\color{black} We use the exact solution of the KdV$\theta$ equation given by
\begin{eqnarray}
&&\eta(T,\xi,\theta) = 2\tilde v \operatorname{sech}^2 \left(\frac{1}{2} \sqrt{6\tilde v} \left[ \xi - \tilde v T - \xi_0 \right] \right),
\label{5}
\end{eqnarray}
where $\xi = rk(\theta) - st$, $T = \epsilon t$, to define the plane-wave sections of a hybrid wave. We match the two plane waves to an arc of a ring wave centred at the origin  at the polar angles $\theta = \pm \varphi$.

The plane-wave sections are defined for $|y| \geq x \tan\varphi$, and we use the general solution of (\ref{GB}),  $k(\theta) = a \cos\theta + b(a) \sin \theta$, and hence $\xi = ax + b(a)y - t$, with $a = \cos \varphi$ and $b(a) = \sin \varphi$.}  In the  $xy$-plane, using the Cartesian velocities (\ref{kdvtvel}),  the initial condition becomes
\begin{eqnarray}
&&\eta(t_0,x,y) = 2\tilde v \operatorname{sech}^2\left(\frac{1}{2} \sqrt{6\tilde v} \left[ ax \pm b(a)y - (1 + \epsilon \tilde v)t_0 - \xi_0 \right] \right), \label{6} \\
&&u(t_0,x,y) = \cos\varphi ~ \eta(t_0,x,y), \quad
v(t_0,x,y) = \pm \sin\varphi ~ \eta(t_0,x,y_0), \label{end6}
\end{eqnarray}
where the upper and lower signs in $\eta$ and the velocity $v$ are for the upper and lower plane-wave sections,  respectively. 

For $|y| < x \tan\varphi$, we take $k(\theta)$ to be the singular solution of (\ref{GB}), {\color{black} i.e. the geometrical envelope of the general solution}, giving $k(\theta) = 1$, and hence $\xi = r - t$. Therefore, in the $xy$-plane, the ring-wave {\color{black} part of the }  initial condition is given by
\begin{eqnarray}
&&\eta(t_0,x,y) = 2\tilde v \operatorname{sech}^2\left(\frac{1}{2} \sqrt{6\tilde v} \left[ \sqrt{x^2 + y^2} - (1 + \epsilon \tilde v)t_0 - {\color{black} r_0} \right] \right), \label{7} \\
&&u(t_0,x,y) = \frac{x}{\sqrt{x^2+y^2}} ~ \eta(t_0,x,y), \quad
v(t_0,x,y) = \frac{y}{\sqrt{x^2+y^2}} ~ \eta(t_0,x,y). \label{end7}
\end{eqnarray}
{\color{black} The initial condition is therefore defined in the entire $xy$-plane, and we match the sections by choosing  the constants $\xi_0 = r_0$, so that at the initial moment of time there is a smooth transition between plane and ring parts at $\theta = \pm \varphi$.}

\begin{figure}
	\centerline{\includegraphics[width=0.9  \linewidth]{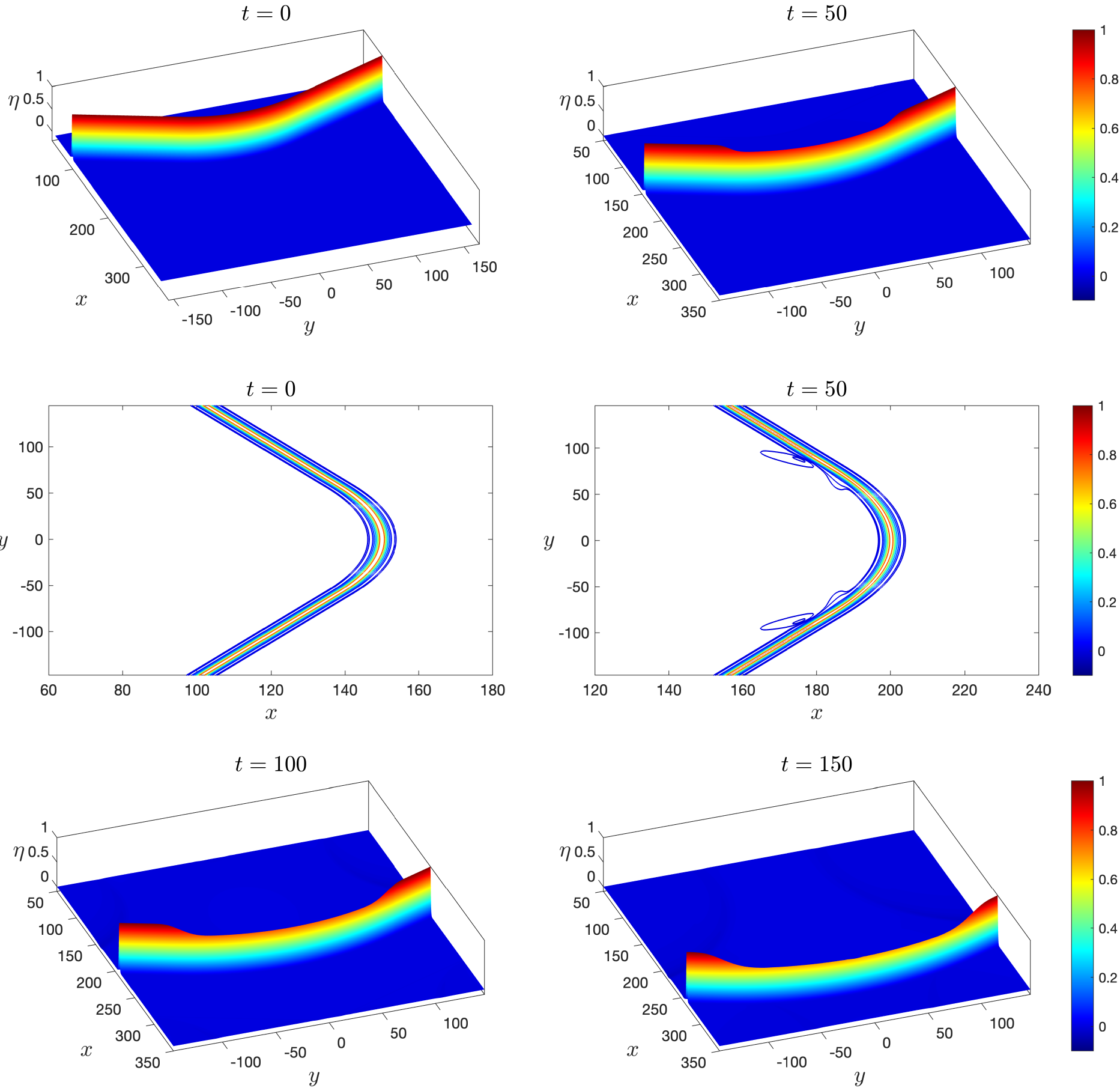}}
	\caption{3D and contour plots of an outward-propagating hybrid wave in the 2D Boussinesq--Peregrine system for the parameter values $\epsilon = 0.01$, $\tilde v = 0.5$, $\varphi = \pm \pi/8$, and $r_0 = 150$ at the times  $t = 0, 50, 100, 150$.}
	\label{Hybrid3DOut}
\end{figure}

We modify the initial conditions to bring them to zero near the boundaries by multiplying them by the double $\tanh$-function {\color{black} $M_x(x) M_y(y)$, where}
\begin{eqnarray}
&&M_x(x) = \frac{1}{2} \left[\tanh \kappa(x - x_{min} - x_{\text{span}}) - \tanh \kappa(x - x_{max} + x_{\text{span}}) \right], \\
&&M_y(y) = \frac{1}{2} \left[\tanh \kappa(y - y_{min} - y_{\text{span}}) - \tanh \kappa(y - y_{max} + y_{\text{span}}) \right],
\end{eqnarray}
%{\color{black} and similarly for $u$ and $v$,} 
{\color{black} to maintain the periodic boundary conditions during our pseudospectral numerical runs.} Typically, we take the parameter values $\kappa = 0.25$, $x_{\text{span}} = (x_{\text{max}} - x_{\text{min}})/20$, and $y_{\text{span}} = (y_{\text{max}} - y_{\text{min}})/20$. Such modification to the amplitude generates cylindrical radiation at the ends of the legs, {\color{black} similar to that described by \citet{YW}, and we ensure that the computational domain is sufficiently large, so that these artefacts do not affect the central region. We solve the 2D Boussinesq--Peregrine system (\ref{Bouss1})--(\ref{Bouss2}) for $t \in [0,150]$, $\epsilon = 0.01$, $\tilde v = 0.5$, $r_0 = 150$, and computation parameters listed in table~\ref{table_hybrid}. The results are shown in} figure \ref{Hybrid3DOut}.

\begin{table}
  \begin{center}
\def~{\hphantom{0}}
  \begin{tabular}{ccccccc}
      Regime        & Direction & $x \times y$			 		 & $N_x$     & $N_y$ & $\Delta_t$ & $\varphi$ \\[5pt]
      2D Boussinesq & Outward   & $[50,350] \times [-250,250]$   & 2400      & 4000  & $10^{-2}$  & $\pi/8$ \\
      2D Boussinesq & Inward    & $[-220,130] \times [-250,250]$ & 2800      & 4000  & $10^{-2}$  & $\pi/4$ \\[10pt]
      Regime        & Direction & $\xi \times Y$			     & $N_{\xi}$ & $N_Y$ & $\Delta_T$ & $\varphi$ \\[5pt]
      KPII          & Outward   & $[0,300] \times [-30,30]$      & 2800      & 4000  & $10^{-3}$  & $\pi/8$ \\
      KPII          & Inward    & $[-200,250] \times [-30,30]$   & 2250      & 1500  & $10^{-3}$  & $\pi/4$
  \end{tabular}
  \caption{Computation parameters for the 2D Boussinesq--Peregrine system and KPII equation simulation of hybrid waves.}
  \label{table_hybrid}
  \end{center}
\end{table}

The initial condition evolves into a wavefront with three distinct sections and a small amount of radiation, not visible here in figure \ref{Hybrid3DOut}. The plane-wave legs propagate with constant amplitude, retaining their shape, and are {\color{black} well  described by the exact soliton solution (\ref{6})--(\ref{end6}), obtained using}  the KdV$\theta$ equation (\ref{A2T}). {\color{black}  The amplitude of the ring-wave section gradually decreases due to the radial divergence, and, provided the wave is sufficiently far from the origin, it is well described  by the cKdV equation (\ref{Bouss_cKdV})} \citep{STCK}. The third section is the transition region between the legs and the central ring wave. It is not described by either the KdV$\theta$ or cKdV equations, but is described by the KPII equation {\color{black} (\ref{KP}),  and, potentially, the modulation theory of \citet{BHM,RMBH} could also be extended to describe this region.} 

{\color{black} This outward-propagating hybrid wave can be described using the KPII equation {\color{black} (\ref{KP})}. To do so, one must rewrite the initial conditions and relevant regions using the KPII variables $T = \epsilon t$, $\xi = x - t$ and $Y = \sqrt{\epsilon} y$}. The plane-wave section {\color{black} ($|y| \geq x\tan\varphi$) is mapped to} $|Y| \geq \sqrt{\epsilon} (\xi + T_0/\epsilon) \tan\varphi$, where {\color{black} the surface elevation (\ref{6}) is given by}
\begin{equation}
\tilde \eta(T_0,\xi,Y) = 2\tilde v \operatorname{sech}^2 \left(\frac{1}{2} \sqrt{6\tilde v} \left[ a \xi + \frac{b(a)}{\sqrt{\epsilon}}Y - \left(\tilde v + \frac{1 - a}{\epsilon} \right) T_0 - r_0 \right] \right). \label{KPplane}
\end{equation}
The ring-wave section {\color{black} ($|y| < x\tan\varphi$) is mapped to} $|Y| < \sqrt{\epsilon} (\xi + T_0/\epsilon) \tan\varphi$, {\color{black} where the surface elevation (\ref{7}) is given by}
\begin{equation}
\tilde \eta(T_0,\xi,Y) = 2\tilde v \operatorname{sech}^2 \left(\frac{1}{2} \sqrt{6\tilde v} \left[ \sqrt{ \left(\xi + \frac{T_0}{\epsilon} \right)^2 + \frac{Y^2}{\epsilon}} - \left(\tilde v + \frac{1}{\epsilon} \right) T_0 - r_0 \right] \right). \label{KPring}
\end{equation}
{\color{black} Initial conditions for the KPII equation must have  zero mass \citep{KSM,BBH}. To abide by this constraint, we modify the initial condition $\tilde\eta$ by adding, for each $Y_i$ in the $Y$-domain, a pedestal of the form
\begin{equation}
p_i = \frac{p_0}{2} \left(\tanh \kappa (\xi - \xi_{\text{min}} - \xi_{\text{span}}) - \tanh \kappa (\xi - \xi_{\text{max}} - \xi_{\text{span}}) \right),  
\end{equation}
where $p_0$ and $\kappa$ are some suitable constants. One can then define the massless initial condition to be
\begin{eqnarray}
&&\eta(T_0,\xi,Y_i) = \tilde\eta(T_0,\xi,Y_i) - p_i(\xi), %\quad \mbox{where} \\
\end{eqnarray}
where
\begin{eqnarray}
&& \int_{\xi_{\text{min}}}^{\xi_{\text{max}}} \tilde\eta(T_0,\xi,Y_i) ~ \mathrm{d}\xi = \int_{\xi_{\text{min}}}^{\xi_{\text{max}}} p_i(\xi) ~ \mathrm{d}\xi.
\end{eqnarray}}
We usually take $\kappa = 0.5$, $\xi_{\text{span}} = (\xi_{\text{max}} - \xi_{\text{min}})/20$. In practice, due to the {\color{black} size}  of the domain, this leads to a pedestal of amplitude less than $\epsilon$, and therefore the {\color{black} effect on numerical solution} is negligible. {\color{black} Further discussion of the numerical solution to the KPII equation is given in Appendix \ref{appB}.}

{\color{black} As the outward-propagating hybrid wave moves away from the origin, the effects of cylindrical divergence are weak}, and the wavefront propagates in a quasi-stationary manner. The description of the central part of the wavefront can therefore be given analytically using the {\color{black} asymptotic solution constructed by \citet{J99}. It is given} for the initial-value problem
\begin{eqnarray}
&& 2 \eta_T + 3\eta\eta_{\xi} + \frac{1}{3} \eta_{\xi\xi\xi} + \frac{\eta}{T} = 0, \quad
\eta\left(T_0,\xi \right) = A \operatorname{sech}^2 \left (\frac{\sqrt{3A}}{2} \xi \right ),
\end{eqnarray}
which is posed  sufficiently far away from the origin, and then the radial-divergence term can be viewed as a perturbation of the KdV equation. The three components of the wave are the primary wave, the shelf and the transition region  leading back to the undisturbed medium. The variables
\begin{eqnarray}
&&\sigma = T_0^{-1}, \quad X = \sigma T, \quad \mathcal{T} = \frac{1}{2}\sqrt{3AX^{-\frac{2}{3}}}, \quad f(X) = \frac{3}{2}A\left(X^{\frac{1}{3}} - 1 \right), \\
&&\Theta = \xi - \sigma^{-1} f(X), \quad \bar{s} = \operatorname{sech}(\mathcal{T}\Theta), \quad \bar{t} = \tanh(\mathcal{T} \Theta),
\end{eqnarray}
are required to construct the three components. The primary wave is given by
\begin{equation}
\hspace{-0.2cm}\textcolor{black}{ \eta_{\text{primary}} = AX^{-\frac{2}{3}}\bar{s}^2 + \frac{2\sigma}{3}\frac{X^{-\frac{2}{3}}}{\sqrt{3A}} \left[-1 + \bar{t} + (3 + 2\mathcal{T}\Theta)\bar{s}^2 - \left(\frac{35}{12} + 3\mathcal{T}\Theta + \mathcal{T}^2\Theta^2 \right)\bar{t}\bar{s}^2 \right], }
\end{equation}
the shelf is described by
\begin{equation}
\textcolor{black}{ \eta_{\text{shelf}} = AX^{-\frac{2}{3}}\bar{s}^2 - \frac{4\sigma}{3} \left(3A + 2 \sigma \xi \right)^{-\frac{1}{2}}, }
\end{equation}
and, finally, the transition region is given in the form
\begin{equation}
\textcolor{black}{ \eta_{\text{transition}} = AX^{-\frac{2}{3}}\bar{s}^2 - \frac{4\sigma}{3} \left(3A \right)^{-\frac{1}{2}} \left(1 - \int_{\xi (2\sigma/X)^{1/3}}^{\infty} \Ai(\xi') ~ \mathrm{d}\xi' \right), }
\end{equation}
where $\Ai$ is the Airy function. The three components can be used to describe the central part of the hybrid wave in $(T,\xi)$-space for suitably small $\sigma$ and, hence, suitably large $T_0$. {\color{black} This asymptotic solution was tested for a pure ring wave by \citet{STCK}, and therefore is applicable in the central part of the hybrid wave, which  is well described by the cKdV equation. Hence, one can describe analytically the three {\color{black} main parts} of the outward-propagating hybrid wave: the two legs and the central region. }

\begin{figure}
	\centerline{\includegraphics[width= 0.9  \linewidth]{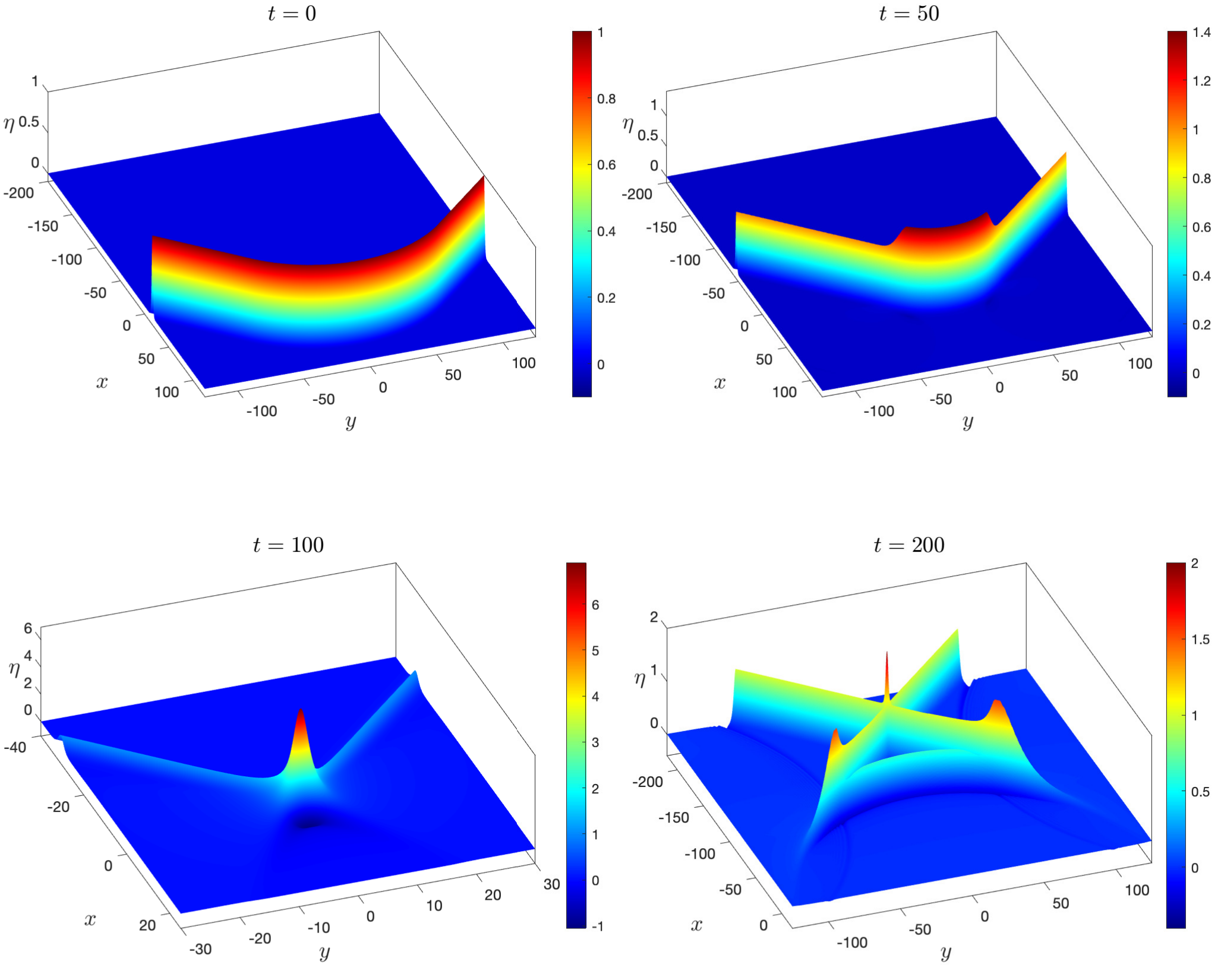}}
	\caption{3D plot of an inward-propagating hybrid wave in the 2D Boussinesq--Peregrine equations for the parameter values $\epsilon = 0.01$, $\tilde v = 0.5$, $\varphi = \pm \pi/4$, and $r_0 = 100$ at the times  $t = 0, 50, 100$, and $150$.}
	\label{Hybrid3DIn}
\end{figure}

\begin{figure}
	\centerline{\includegraphics[width=0.8 \linewidth]{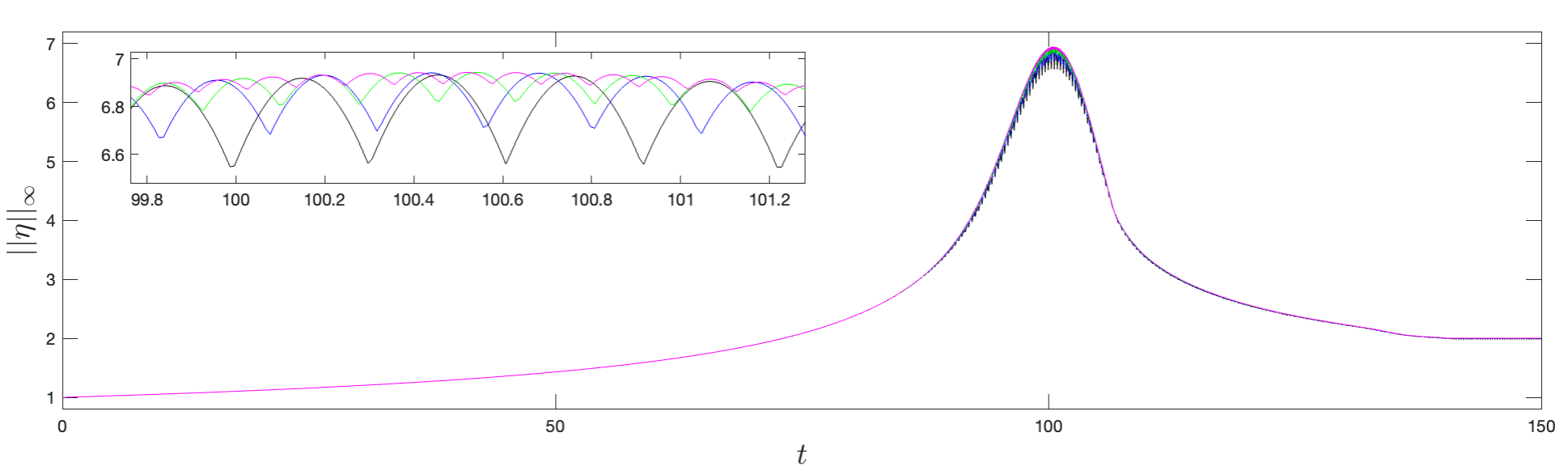}}
	\caption{The $L^\infty$ norm of $\eta$ from $t=0$ to $t = 150$ for the inward-propagating hybrid wave with $\epsilon = 0.01$ and $\varphi = \pi/4$ where $\Delta_x = \Delta_y = 0.35$ (black),  $\Delta_x = \Delta_y = 0.275$ (blue),  $\Delta_x = \Delta_y = 0.2$ (green), and  $\Delta_x = \Delta_y = 0.125$ (magenta).}
	\label{L_infty_norm}
\end{figure}

\begin{figure}
	\centerline{\includegraphics[width=0.8 \linewidth]{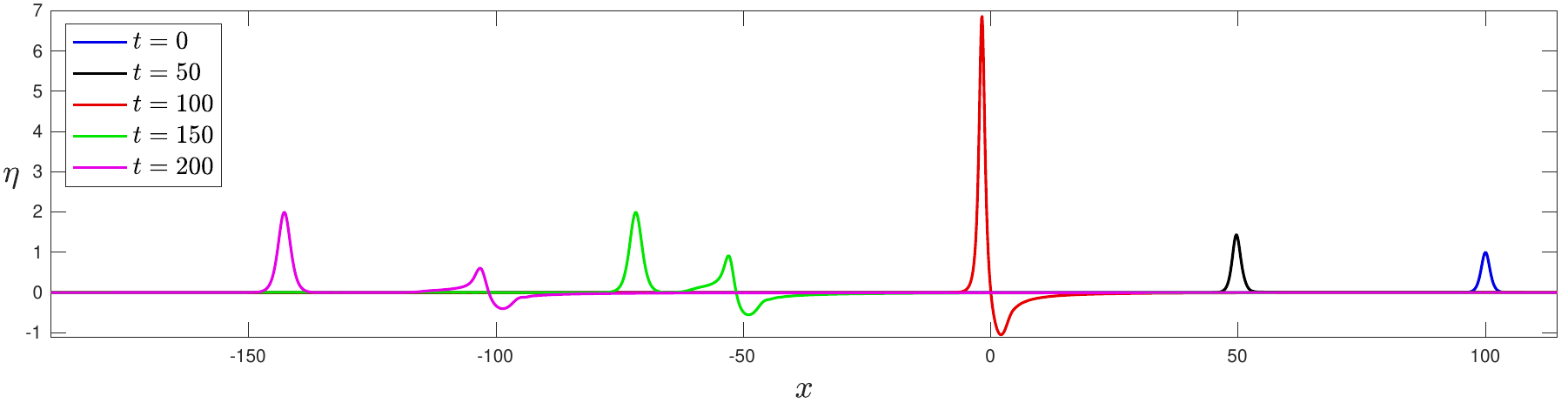}}
	\caption{Cross-section plot along $x = 0$ of the domain plotted at 5 times from $t = 0$ to $t = 200$.}
	\label{hybrid_y0}
\end{figure}

We now model the inward-propagating hybrid wave. The wave is generated by the same set of initial conditions, {\color{black} (\ref{6})--(\ref{end7})}, in the same regions. However, the velocities are multiplied by $-1$, to reverse the direction of propagation, and similarly brought to zero near the boundaries. {\color{black} We solve the 2D Boussinesq--Peregrine system (\ref{Bouss1})--(\ref{Bouss2})  for $t \in [0,150]$, $\epsilon = 0.01$, $\tilde v = 0.5$, $r_0 = 100$, and the computation parameters} listed in table~\ref{table_hybrid}. The results are shown in figures \ref{Hybrid3DIn}, \ref{L_infty_norm} and \ref{hybrid_y0}.

The intermediate evolution of the wavefront is dominated by the ring-wave section focusing in on itself, increasing the amplitude dramatically, as shown by the $L^{\infty}$ norm in figure \ref{L_infty_norm}. However, this is not a finite-time blow-up of the numerical solution, since, as the resolution of the mesh is increased, the amplitude converges to a consistent value {\color{black} for the same} $t$-value. The converging wavefront generates a large rogue wave {\color{black} (lump)}, which subsequently disintegrates into three pieces and eventually generates a stable ``X-type'' formation, similar to that observed within the scope of the KPII equation in figure 1 of \citet{AB} and studied in \citet{CK09,CK,CLM}. {\color{black} There also exists} an arc  of opposite curvature to the initial condition connecting two lump-like structures on the legs behind the X. Previous studies have focused on ``V-type'' initial data for the KPII equation rather than our hybrid wave formation. These studies have shown similar ``X-type''  structures obtained numerically without a noticeable phase shift, see, for example,  \citet{CK09,RMBH,RHB}. A weaker arc of opposite curvature to the initial condition connecting the two legs was observed  by \citet{RMBH,YGJW} in the KPII equation and from the cross-section in figure \ref{hybrid_y0} has {\color{black} the shape similar} to the plane wave derivative. {\color{black} The effect of curvature on the amplitude of the generated waves was noted in \citet{PTLO}.}  To the best of our knowledge, the overall behaviour of the inward-propagating hybrid wave, including the generation of the large rogue wave and its disintegration into the three pieces, {\color{black} with the subsequent formation of the X-wave},  has not been reported in any previous studies.  Although the KPII equation does not support the exact lump solutions, unlike the KPI equation, our study shows that lumps, {\color{black} understood as localised 2D waves}, can exist as transient {\color{black}(emerging and then disappearing)} states in the evolution of curved solitons, contributing to the mechanisms of generation of rogue waves (see, for example, \citealt{KPS, ORBMA, SP, HC, NTK} and references therein). 

\begin{figure}
	\centerline{\includegraphics[width= 0.9 \linewidth]{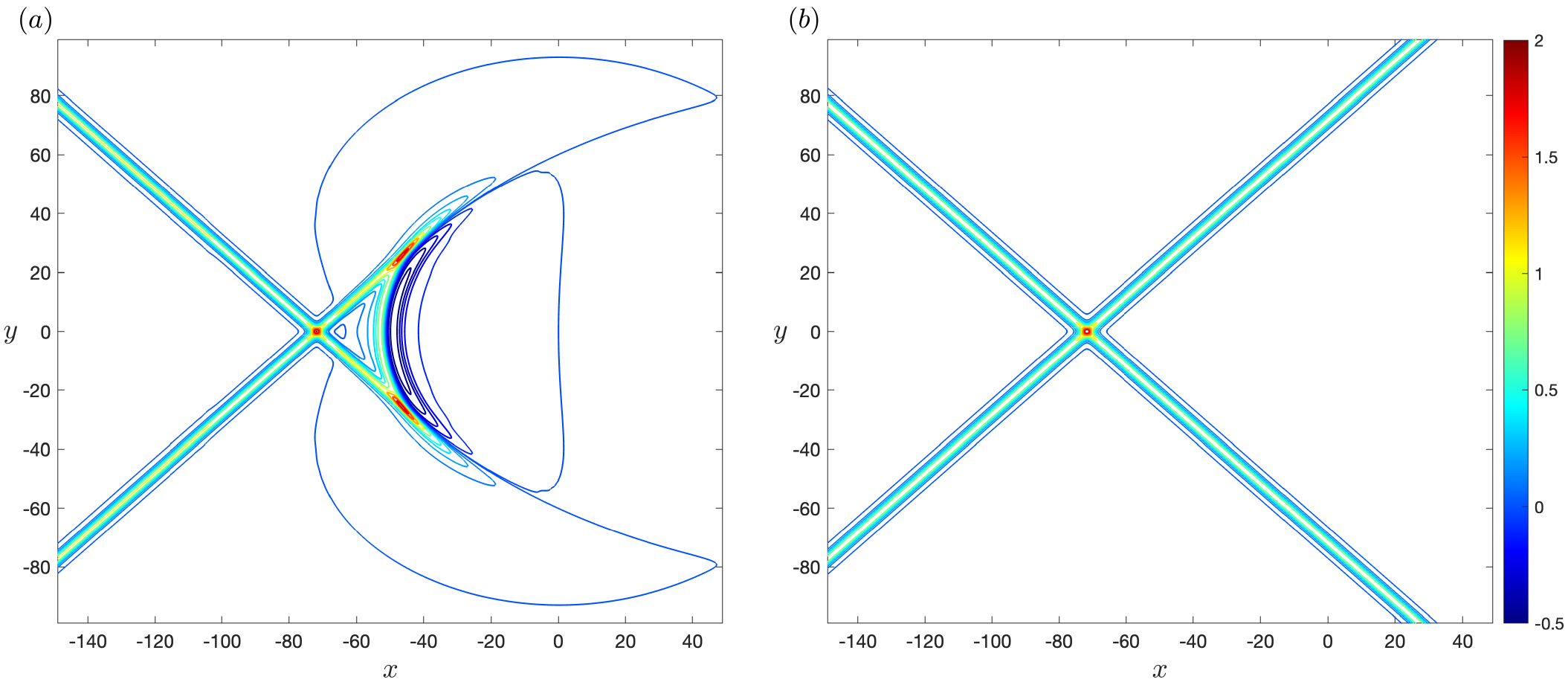}}
	\caption{Contour plots of \textit{(a)} the inward-propagating hybrid wave at $t = 150$ from figure \ref{Hybrid3DIn} and \textit{(b)} the parameter-matched two-soliton solution of the KPII equation.}
	\label{KP_X_comp}
\end{figure}

{\color{black} It turns out that the ``X-type'' wave formed at the end can be  described by the solution of the KPII equation discussed, in a different context, by  \citet{AB}:
\begin{equation}
\eta = \frac{4}{3} \frac{\partial^2}{\partial\xi^2} \log \left( F_2 \right),
\end{equation}
where
\begin{eqnarray}
&&F_2 = 1 + e^{\eta_1} + e^{\eta_2} + e^{\eta_1 + \eta_2 + A_{12}}, \\
&&\eta_i = 6k_i \left( \xi + 6P_iY - 6\left( k_i^2 + 3 P_i^2 \right) T \right) + c_i, \quad i = 1,2, \\
&&e^{A_{12}} = \frac{(k_1 - k_2)^2 - (P_1 - P_2)^2}{(k_1 + k_2)^2 - (P_1 - P_2)^2},
\end{eqnarray}
such that $k_i$ and $P_i$ are constants determining the propagation angle and $c_i$ is an arbitrary constant shifting the wavefront. For our numerical run,  the crossing angle is $\pi/4$, and good agreement is obtained for the parameter values $6k_1 = 6k_2 = 1.1314$, $6P_1 = -6P_2 = \epsilon^{-1/2}$ and $c_1 = c_2 = 71.75k_1$. }The value of $e^{A_{12}}$ is indicative of the behaviour of the wavefront where the two solitons cross. For the case seen here, $e^{A_{12}} = 1.013$, implying an ``X-type'' interaction with a very short stem between the legs, as seen from the evolved state of the inward-propagating hybrid wavefront. Indeed, there is no observable phase shift and the comparison between the KPII two-soliton solution and the inward-propagating hybrid wave is given in figure \ref{KP_X_comp}. The amplitude of the two-soliton cross depends on the propagation angle, and a discussion on how to obtain the maximum of the cross is discussed by \citet{CLM}.

\begin{figure}
	\centerline{\includegraphics[width=  \linewidth]{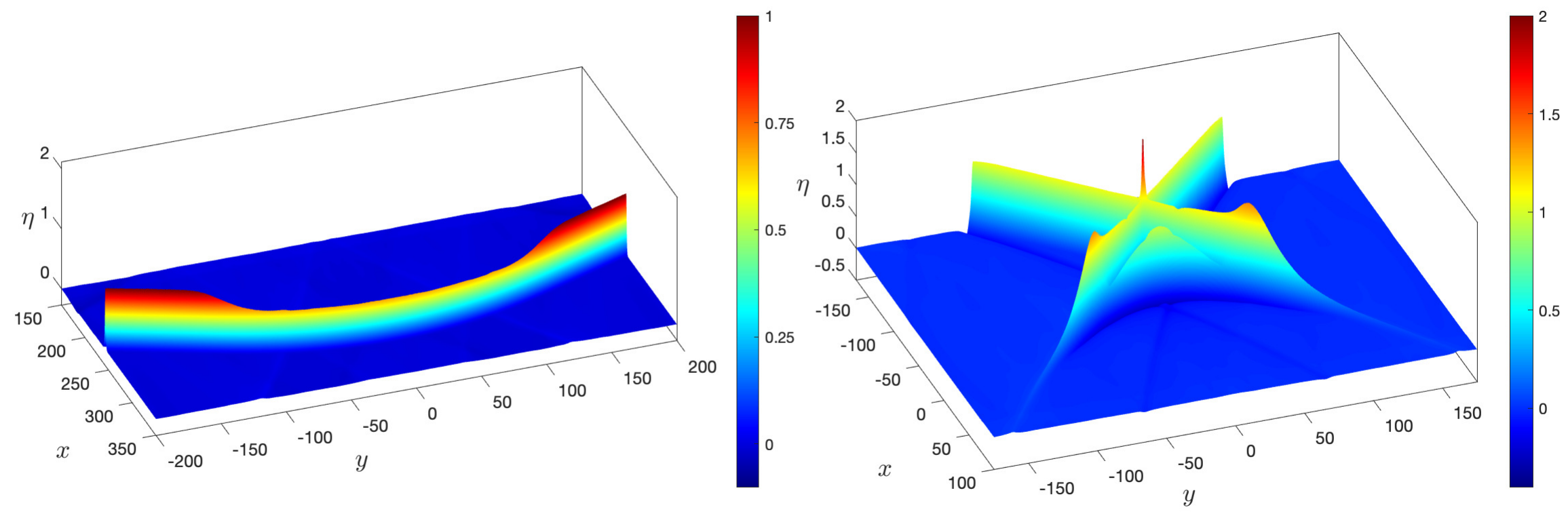}}
	\caption{3D plots of the numerical solution to the KPII equation for both the outward- and inward-propagating hybrid waves solved for $T \in [0,1.5]$, $\epsilon = 0.01$ and $\tilde v=0.5$ where for \textit{(a)} $\varphi = \pi/8$ and $\xi_0 = 150$, and for \textit{(b)} $\varphi = \pi/4$ and $\xi_0 = 100$.}
	\label{KP_hybrid_in}
\end{figure}

To model the entire process of the inward-propagating hybrid wave seen in figure \ref{Hybrid3DIn} with the KPII equation, we must first reverse the direction of propagation. Therefore, for $T$ reducing, we map $t \rightarrow -t$ and hence $\partial/\partial T = - \partial/\partial T$  and $\xi \rightarrow \zeta = x + t$, which yields the inward-propagating KPII equation
\begin{equation}
\left( 2\eta_T - 3\eta\eta_{\zeta} - \frac{1}{3} \eta_{\zeta\zeta\zeta} \right)_{\zeta} - \eta_{YY} = 0. \label{KPin}
\end{equation}
The initial condition {\color{black} for the plane-wave section ($|y| \geq x \tan\varphi$)} is defined when $|Y| \geq \sqrt{\epsilon} (\zeta - T_0/\epsilon) \tan\varphi$ and is given by
\begin{equation}
\tilde \eta(T_0,\zeta,Y) = 2\tilde v \operatorname{sech}^2 \left(\frac{1}{2} \sqrt{6\tilde v} \left[ a \zeta + \frac{b(a)}{\sqrt{\epsilon}}Y - \left(\tilde v + \frac{1 + a}{\epsilon} \right) T_0  - \xi_0 \right] \right), \label{KPplanein}
\end{equation}
and the ring-wave section {\color{black} ($|y| < x \tan\varphi$)}, defined when $|Y| < \sqrt{\epsilon} (\zeta - T_0/\epsilon) \tan\varphi$, is given by
\begin{equation}
\tilde \eta(T_0,\zeta,Y) = 2\tilde v \operatorname{sech}^2 \left(\frac{1}{2} \sqrt{6\tilde v} \left[ \sqrt{ \left(\zeta - \frac{T_0}{\epsilon} \right)^2 + \frac{Y^2}{\epsilon}} - \left( \tilde v + \frac{1}{\epsilon} \right) T_0 - \xi_0 \right] \right), \label{KPringin}
\end{equation}
where, before solving, the mass of the solution should be removed to satisfy the zero-mass constraint of the KPII equation as before, such that $\eta = \tilde \eta - p(\xi)$ for the pedestal $p(\xi)$. Solving (\ref{KP}) and (\ref{KPin}) with the initial data of (\ref{KPplane}), (\ref{KPring}), and (\ref{KPplanein}), (\ref{KPringin}), respectively, for the parameter values given to match figures \ref{Hybrid3DOut} and \ref{Hybrid3DIn}, we obtain the results that are shown in figure \ref{KP_hybrid_in}.

{\color{black} We see that the computations give all of the same structures as those observed in the 2D Boussinesq--Peregrine system, up to the accuracy of the reduced model.  Notably, the transient lump-like structures in the middle and at the ends of the legs are present, as in figure \ref{Hybrid3DIn}, despite the fact that the KPII equation for weak surface tension used here does not possess lump soliton solutions.} These transient lumps constantly decrease in amplitude, converting their energy to extend the legs behind the X. We also note that some radiation escapes and re-enters  the periodic domain of the numerical solver, causing some small defects. The inclusion of a sponge layer can remove these artefacts.

{\color{black} Finally, in this section we model the evolution of an inward-propagating perturbed hybrid wave. We choose to apply a periodic perturbation to the ring-wave section, altering (\ref{7})--(\ref{end7}) but keeping (\ref{6})--(\ref{end6}) unchanged. We again take the velocities to be negative for inward propagation. We perturb the arc of a ring wave in the region $|y| < x \tan\varphi$ to be given by
\begin{eqnarray}
&&\eta(t_0,x,y) = \left( \frac 32 + \alpha \cos \left( \beta \arctan \frac yx \right)  \right) \eta_0(t_0,x,y), \label{perteta1} \\
&&u(t_0,x,y)    = - \frac{x}{\sqrt{x^2+y^2}} ~ \eta(t_0,x,y), \quad
v(t_0,x,y)    = - \frac{y}{\sqrt{x^2+y^2}} ~ \eta(t_0,x,y). \label{perteta3}
\end{eqnarray}
Solving the 2D Boussinesq--Peregrine system (\ref{Bouss1})--(\ref{Bouss2})  for the perturbed initial condition given by (\ref{6})--(\ref{end6}) and (\ref{perteta1})--(\ref{perteta3}) for $t \in [0,150]$, $\epsilon = 0.01$, $\tilde v = 0.5$, $r_0 = 100$, $\alpha = 0.5$, $\beta = 12$, and the computation parameters listed in table~\ref{table_hybrid},  yields the results shown in figure \ref{perturbed_hybrid}.

\begin{figure}
	\centerline{\includegraphics[width=\linewidth]{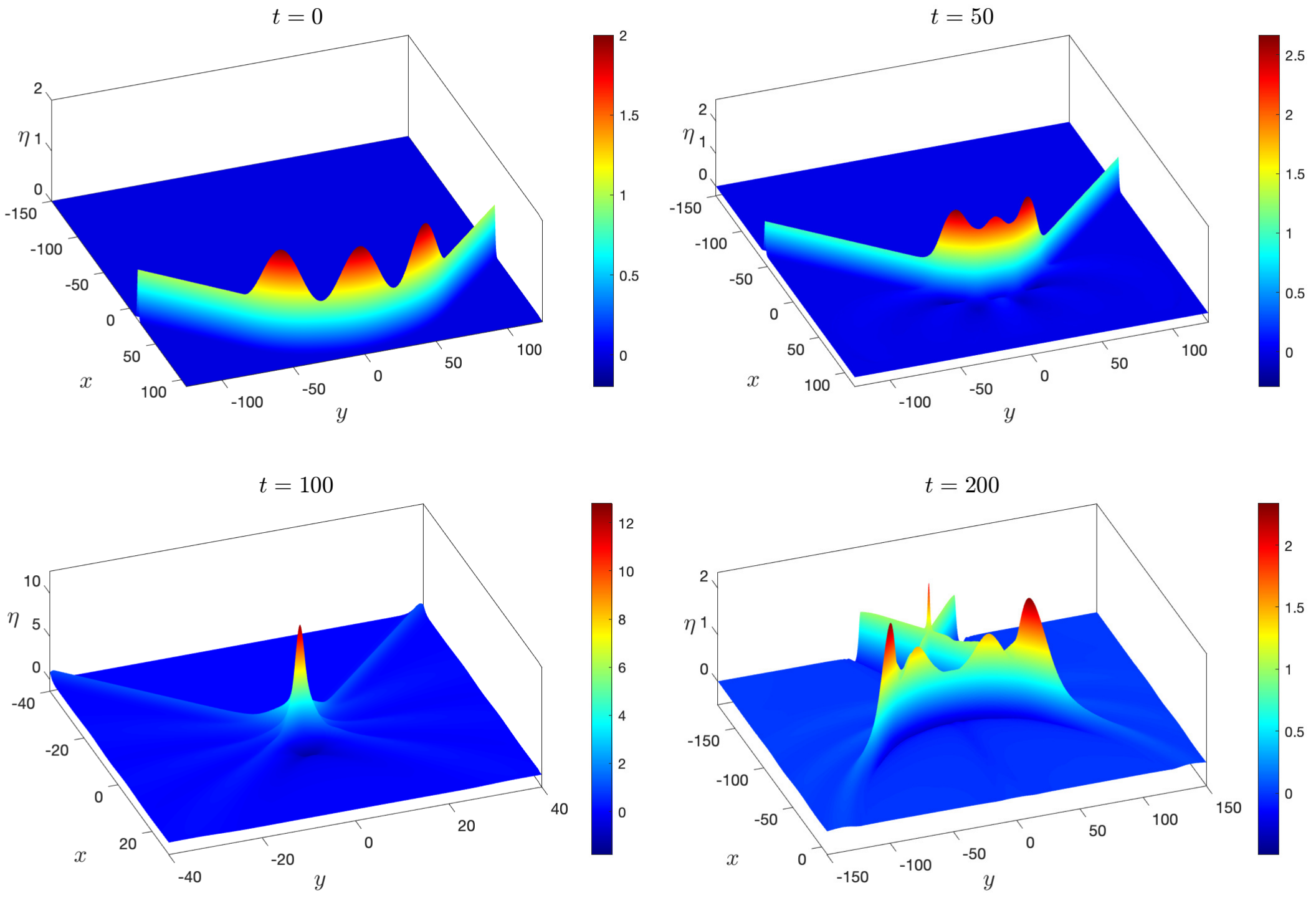}}
	\caption{{\color{black} 3D plots of a perturbed inward-propagating hybrid wave in the 2D Boussinesq--Peregrine system for the parameter values $\epsilon = 0.01$, $\tilde v = 0.5$, $\varphi = \pm \pi/4$, $r_0 = 100$, $\alpha = 0.5$, and $\beta = 12$, at the times  $t = 0, 50, 100, 150$.}}
	\label{perturbed_hybrid}
\end{figure}

The evolution of the inward-propagating perturbed hybrid wave is remarkably similar to that of the unperturbed case, despite the large amplitude and wavelength of the perturbation applied. The perturbed hybrid wave also generates a large rogue wave (larger than in the unperturbed case), and still eventually forms a stable ``X-type'' cross. The initial radiation behind the ring-wave section is qualitatively similar to  that observed in the ring-wave study in Section 4. The main difference is that the transient rogue wave disintegrates into more pieces than in the pure hybrid wave evolution. 

Finally, we note that the $2+1$-dimensional cKdV-type equation (\ref{A}) can be used to initiate hybrid waves in the general case of a stratified fluid with a shear flow (see \citealt{HKG}). 

}

% ---------- ---------- ---------- ---------- %
\section{Discussion}

{\color{black} In this paper, we investigated the non-stationary two-dimensional evolution of perturbed plane, ring, and hybrid waves within the scope of the 2D Boussinesq--Peregrine system \citep{P} for surface waves in a homogeneous fluid, while simultaneously discussing how to extend the theoretical constructions to the general case of a density stratified fluid with a parallel shear flow. In particular, for the plane-wave evolution, we derived a new amplitude equation, the KdV$\theta$ equation (\ref{A2}), and showed that this equation can be mapped to the usual KdV equation, but with a $\theta$-dependent initial condition. We showed that this result holds for any stratification, and any parallel shear flow, using the 2+1-dimensional cKdV-type equation derived by \citet{KZ}. We then applied the KdV$\theta$ equation to describe the evolution of a line soliton subject to a long transverse perturbation. We obtained explicit analytical predictions for the evolution using the Inverse Scattering Transform \citep{GGKM}, and compared the performance of the reduced model with the results of direct numerical simulations for the 2D Boussinesq--Peregrine system in order to clarify the range of validity of the model. We showed that the KdV$\theta$ equation can be used to describe the intermediate evolution of sufficiently long transverse perturbations of line solitons (up to the moment when the perturbation splits and starts propagating sideways along the wavefront). Up to this point the KdV$\theta$ equation accurately describes the behaviour of the solution. Hence, when applicable, the KdV$\theta$ equation allows one to consider the initial-value problem, which serves as a useful addition to the KPII equation (e.g. \cite{AS}). Recently, transverse amplitude and phase perturbations to line solitons were modelled numerically in the Serre--Green--Naghdi regime by \citet{GK}, and the results are qualitatively similar to plane waves propagating over localised bottom topography (see \citealt{BBBD, GS}). The new equation could  thus find useful applications in these settings.

Next, we modelled the non-axisymmetric evolution of the perturbed ring waves, extending the earlier studies of concentric waves by \citet{CW, STCK}. We considered both localised and periodic perturbations, and modelled the long-time evolution of both outward- and inward-propagating waves. The results related to the cKdV equation \citep{M, J90, J99} were used to formulate the basic unperturbed initial conditions, but naturally, the cKdV equation could not be used to model the subsequent non-axisymmetric evolution of the perturbed initial conditions, and we used the direct numerical simulations with the parent system instead. In all our runs, the outward-propagating ring waves were stable, while stability of the inward-propagating waves turned out to be related to the characteristic length of the perturbation. In particular, we illustrated the qualitative theoretical predictions made by \citet{OS,SP,P91} and showed the difference between the evolution of ring waves with periodic perturbations of differing wavelength, with the waves of smaller/greater wavelength being stable/unstable, respectively. Deriving an explicit formula for the critical wavelength of the perturbation is an interesting open problem.

Finally, we modelled the evolution of hybrid waves, initially consisting of a part of a ring wave and two tangent plane waves. To the best of our knowledge, this is the first systematic study of such waves. We used the results related to the KdV$\theta$, cKdV and KPII equations in order to initiate our two-dimensional simulations, and to describe the key features of the evolution. The outward-propagating hybrid waves were found to be stable, and, when generated sufficiently far away from the origin (in the far-field), propagated as quasi-stationary waves, with the central part slowly decreasing in amplitude due to cylindrical divergence. We note that for the stable cases, qualitatively similar propagation has been observed  from a ``V-type'' (bent soliton) initial conditions by \citet{CK09,YGJW, RMBH, RHB, BBH}. Therefore, it is likely that the key features of the outward-propagating case may be described by Whitham modulation theory (see \citealt{BHM,HLW} and references therein).

The inward-propagating hybrid waves, however, turned out to be unstable, which confirmed the theoretical prediction made, using the ray theory, by \citet{OS}. We showed that the developed stage of this  instability consisted of three main stages:  the generation of a large rogue wave (lump),  its subsequent disintegration  into several pieces, and  the formation of the ``X-wave'' known from the theory of the KPII equation. We found the parameters of the ``X-wave'' matching our results. Moreover, the key stages of this instability scenario did not change when we subjected the central part to a strong periodic perturbation. Finally, we comment that it would  be interesting to extend this work to the case of stratified fluids with a background shear flow (see \citealt{KZ, HKG, TABK}), and to waves of moderate amplitude, which so far have been studied using the extended KdV-type models only in the plane and concentric wave settings (see \citealt{STCK, STCK2} and references therein).}

% ---------- ---------- ---------- ---------- %
%\backsection[Supplementary data]{\label{SupMat}Supplementary material and movies are available at \\https://doi.org/10.1017/jfm.2019...}

\backsection[Acknowledgements]{Karima Khusnutdinova is grateful to Artur Sergeev and Victor Shrira for useful discussions. Benjamin Martin is grateful to Loughborough University for a research studentship.}

\backsection[Funding]{This research received no specific grant from any funding agency, commercial or not-for-profit sectors.}

\backsection[Declaration of interests]{The authors report no conflict of interest.}

%\backsection[Data availability statement]{The data that support the findings of this study are openly available in [repository name] at http://doi.org/[doi], reference number [reference number]. See JFM's \href{https://www.cambridge.org/core/journals/journal-of-fluid-mechanics/information/journal-policies/research-transparency}{research transparency policy} for more information}

\backsection[Author ORCIDs]{
B. Martin, https://orcid.org/0009-0005-6197-8752, \\
D. Tseluiko, https://orcid.org/0000-0002-7676-7641, \\
K. R. Khusnutdinova, https://orcid.org/0000-0002-9201-9694}

%\backsection[Author contributions]{Authors may include details of the contributions made by each author to the manuscript'}

% ---------- ---------- ---------- ---------- %
\appendix

% ---------- ---------- ---------- ---------- %

\section{Derivation of approximate conservation laws of the 2D Boussinesq--Peregrine system}\label{appA}

We calculate the conserved quantities of the 2D Boussinesq--Peregrine system in the rectangular domain, {\color{black} $D= \{(x, y)| x_1 \le x \le x_2, y_1 \le y \le y_2\}$, where we assume periodic boundary conditions. The derivation of the 2D Boussinesq--Peregrine system  is based on the balance $\delta^2 = O(\epsilon)$ \citep{P}, and so one must keep equivalent orders in the conserved form, up to the accuracy of the derivation.
}

The conservation of mass in the system is found directly from (\ref{Bouss1}). By taking the double integral of (\ref{Bouss1}) over $D$ we obtain
\begin{equation}
\frac{\mathrm{d}}{\mathrm{d} t} \iint_D  \eta ~ \mathrm{d}x \mathrm{d}y = - \iint_D  \left[ ((1+\epsilon \eta)u)_x + ((1 + \epsilon \eta)v)_y \right] ~ \mathrm{d}x \mathrm{d}y, \label{mass1}
\end{equation}
where $u$ and $v$ are the Cartesian components of the velocity $\mathbf{u}$. From Green's theorem (\ref{mass1}) can be rewritten as the contour integral around the rectangular domain yielding
\begin{eqnarray}
&& \frac{\mathrm{d}}{\mathrm{d} t} \iint_D  \eta ~ \mathrm{d}x \mathrm{d}y = \oint_C \left[ (1+\epsilon \eta) u ~ 
 \mathrm{d}x - (1 + \epsilon \eta) v ~ \mathrm{d} y \right] \\
 &&= \int_{x_1}^{x_2} (1+\epsilon \eta) v \vert_{y=y_1} ~ \mathrm{d}x + \int_{x_2}^{x_1} (1+\epsilon \eta) v \vert_{y=y_2} ~ \mathrm{d}x  \nonumber \\
 %&&\quad 
 &&- \int_{y_1}^{y_2} (1+\epsilon \eta) u \vert_{x=x_2} ~ \mathrm{d}y - \int_{y_2}^{y_1} (1+\epsilon \eta) u \vert_{x=x_1} ~ \mathrm{d}y 
 %\\
 = 0
\end{eqnarray}
due to periodicity of the problem.

The approximate momentum of the system, a vector quantity, is given by the double integral of $(1+\epsilon\eta)\mathbf{u}$. To obtain an expression for this quantity, we apply the operator $1 + \frac{\delta^2}{3}\nabla (\nabla \cdot\,\, )$ to (\ref{Bouss2}) and multiply by $1 + \epsilon \eta$, which after rearranging yields
\begin{eqnarray}
(1 + \epsilon \eta) u_t &&= -\epsilon(uu_x + vu_y) - (1 + \epsilon \eta) \eta_x + \frac{\delta^2}{3} (\eta_{xxx} + \eta_{xxy}) + O(\epsilon^2, \epsilon\delta^2,\delta^4), \label{momentum1} \\
(1 + \epsilon \eta) v_t &&= -\epsilon(uv_x + vv_y) - (1 + \epsilon \eta) \eta_y + \frac{\delta^2}{3} (\eta_{xyy} + \eta_{yyy}) + O(\epsilon^2, \epsilon\delta^2,\delta^4), \label{momentum2}
\end{eqnarray}
and we also take $\epsilon \mathbf{u}$ multiplied by (\ref{Bouss1}) to give
\begin{eqnarray}
\epsilon u \eta_t &&= - \epsilon(uu_x + uv_y) + O(\epsilon^2), \label{momentum3} \\
\epsilon v \eta_t &&= - \epsilon(vu_x + vv_y) + O(\epsilon^2). \label{momentum4}
\end{eqnarray}
%It is noted that in the derivation of the 2D Boussinesq--Peregrine equations $\delta^2 = O(\epsilon)$ and so one must keep equivalent orders in the conserved form.
% consistent with those truncated. 
Taking the sum of (\ref{momentum1}) and (\ref{momentum3}), and (\ref{momentum2}) and (\ref{momentum4}), respectively, and taking the double integral over the domain we can write the result in the form of a conservation law as
\begin{eqnarray}
&& \frac{\mathrm{d}}{\mathrm{d} t} \iint_D  (1+ \epsilon \eta) u ~ \mathrm{d}x \mathrm{d}y = \iint_D \biggl( \left[ \frac{\delta^2}{3} (\eta_{xx} + \eta_{xy}) - \eta - \epsilon \left(u^2 + \frac{1}{2}\eta^2 \right) \right]_x \nonumber \\
&&- \epsilon \left[ uv \right]_y \biggr) + O(\epsilon^2, \epsilon \delta^2, \delta^4), \label{momentum5} \\
&& \frac{\mathrm{d}}{\mathrm{d} t} \iint_D  (1+ \epsilon \eta) v ~ \mathrm{d}x \mathrm{d}y = \iint_D \biggl( \left[ \frac{\delta^2}{3} (\eta_{xy} + \eta_{yy}) - \eta - \epsilon \left( v^2 + \frac{1}{2}\eta^2 \right) \right]_y \nonumber \\
&& - \epsilon \left[ uv \right]_x \biggr) + O(\epsilon^2, \epsilon \delta^2, \delta^4). \label{momentum6}
\end{eqnarray}
We can then rewrite (\ref{momentum5}) and (\ref{momentum6}) as contour integrals via Green's theorem as
\begin{eqnarray}
&&  \frac{\mathrm{d}}{\mathrm{d} t} \iint_D  (1+ \epsilon \eta) u ~ \mathrm{d}x \mathrm{d}y = \oint_C \biggl[ \left\{ \frac{\delta^2}{3} (\eta_{xx} + \eta_{xy}) - \eta - \epsilon \left(u^2 + \frac{1}{2}\eta^2 \right) \right\} ~ \mathrm{d}y \nonumber \\
&& +\ \epsilon uv ~ \mathrm{d}x \biggr] + O(\epsilon^2, \epsilon \delta^2, \delta^4) \\
&&= \int_{y_1}^{y_2} \left[ \frac{\delta^2}{3} (\eta_{xx} + \eta_{xy}) - \eta - \epsilon \left(u^2 + \frac{1}{2}\eta^2 \right) \right]_{x = x_2} ~ \mathrm{d}y \nonumber \\
&& + \int_{y_2}^{y_1} \left[ \frac{\delta^2}{3} (\eta_{xx} + \eta_{xy}) - \eta - \epsilon \left(u^2 + \frac{1}{2}\eta^2 \right) \right]_{x = x_1} ~ \mathrm{d}y \nonumber \\
&& + \int_{x_1}^{x_2} uv \vert_{y=y_1} ~ \mathrm{d}x + \int_{x_2}^{x_1} uv \vert_{y=y_2} ~ \mathrm{d}x  + O(\epsilon^2, \epsilon \delta^2, \delta^4) \\
&&= O(\epsilon^2, \epsilon \delta^2, \delta^4),
\end{eqnarray}
\begin{eqnarray}
&&  \frac{\mathrm{d}}{\mathrm{d} t} \iint_D  (1+ \epsilon \eta) v ~ \mathrm{d}x \mathrm{d}y = \oint_C \biggl[ \left\{ \eta + \epsilon \left(v^2 + \frac{1}{2}\eta^2 \right) - \frac{\delta^2}{3} (\eta_{xy} + \eta_{yy}) \right\} ~ \mathrm{d}y \nonumber \\
&& - \epsilon uv ~ \mathrm{d}x \biggr] + O(\epsilon^2, \epsilon \delta^2, \delta^4) \\
&&= \int_{x_1}^{x_2} \left[ \eta + \epsilon \left(v^2 + \frac{1}{2}\eta^2 \right) - \frac{\delta^2}{3} (\eta_{xy} + \eta_{yy}) \right]_{y = y_1} ~ \mathrm{d}x \nonumber \\
&&+ \int_{x_2}^{x_1} \left[ \eta + \epsilon \left(v^2 + \frac{1}{2}\eta^2 \right) - \frac{\delta^2}{3} (\eta_{xy} + \eta_{yy}) \right]_{y = y_2} ~ \mathrm{d}x \nonumber \\
&& - \int_{y_1}^{y_2} uv \vert_{x=x_2} ~ \mathrm{d}y - \int_{y_2}^{y_1} uv \vert_{x=x_2} ~ \mathrm{d}y 
+ O(\epsilon^2, \epsilon \delta^2, \delta^4) \\
&&= O(\epsilon^2, \epsilon \delta^2, \delta^4).
\end{eqnarray}
Hence, momentum is conserved in both the $x$- and $y$-directions to the second order of the expansions used to derive the 2D Boussinesq--Peregrine system.

The approximate energy of the system is given by the double integral of $\frac{1}{2}(u^2 + v^2 + \eta^2)$. It is therefore necessary to multiply (\ref{Bouss1}) by $\eta$ and (\ref{Bouss2}) by $\mathbf{u}$ and combine the three subsequent equations. Taking the time derivative of this gives
\begin{equation}
 \frac{\mathrm{d}}{\mathrm{d} t} \iint_D  \frac{1}{2}\left( u^2 + v^2 + \eta^2 \right) ~ \mathrm{d}x \mathrm{d}y = - \iint_D  \left[ \left(u \eta \right)_x + \left(v \eta \right)_y \right] ~ \mathrm{d}x \mathrm{d}y + O(\epsilon,\delta^2),
\end{equation}
which we again rewrite as a contour integral around the rectangular computational domain such that
\begin{eqnarray}
&& \frac{\mathrm{d}}{\mathrm{d} t} \iint_D  \frac{1}{2}\left( u^2 + v^2 + \eta^2 \right) ~ \mathrm{d}x \mathrm{d}y = \oint_C \left[ v\eta ~ \mathrm{d}x - u \eta ~ \mathrm{d} y \right] + O(\epsilon,\delta^2) \\
&&= \int_{x_1}^{x_2} v \eta \vert_{y=y_1} ~ \mathrm{d}x + \int_{x_2}^{x_1} v \eta \vert_{y=y_2} ~ \mathrm{d}x \nonumber \\
&&- \int_{y_1}^{y_2} u \eta \vert_{x=x_2} ~ \mathrm{d}y - \int_{y_2}^{y_1} u \eta \vert_{x=x_1} ~ \mathrm{d}y 
%+ O(\epsilon,\delta^2) \\
= O(\epsilon,\delta^2),
\end{eqnarray}
and, hence, the energy is conserved to the first order of the expansions used to derive the 2D Boussinesq--Peregrine system.

\begin{figure}
	\centerline{\includegraphics[width=0.8 \linewidth]{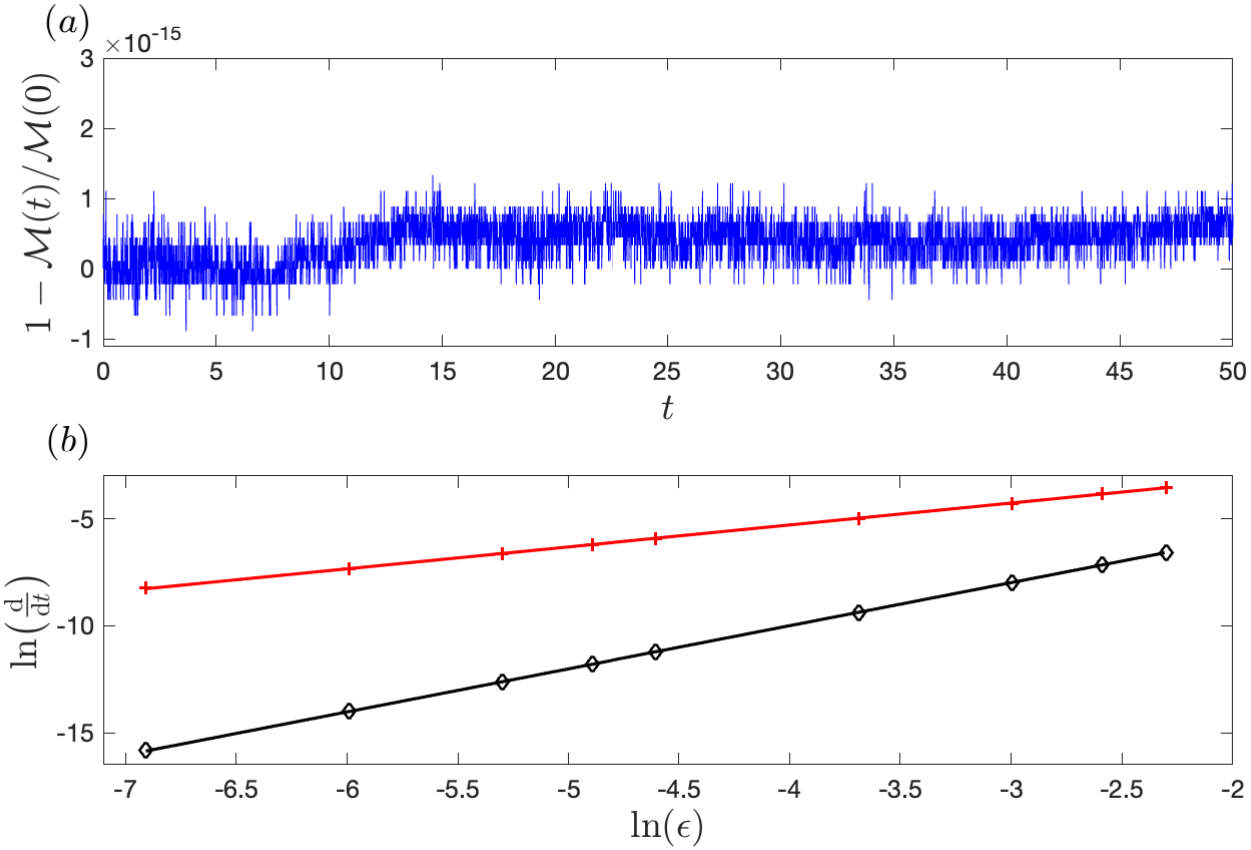}}
	\caption{Conservation of mass, momentum, and energy for the line-soliton initial condition: (\textit{a}) conservation of the mass for $\epsilon = 0.01$ and (\textit{b})  logarithms of the time derivatives at the final computation time $t = 50$ of momentum (red) and energy (black) against the  logarithm of  $\epsilon$.  The  behaviour for energy has a gradient of $1.025$ and the one for momentum a gradient of $2.015$. }
	\label{soliton_quantities}
\end{figure}

{\color{black} The conservation laws are used to control the quality of our numerical simulations. We illustrate this with an example related to the modelling of the line-soliton initial condition}
\begin{eqnarray}
\eta(0,x,y) &&= 2\tilde v \operatorname{sech}^2 \left(\frac{1}{2} \sqrt{6\tilde v} \left[ x - x_0 \right] \right), \\
u(0,x,y) &&= \eta(0,x,y), \\
v(0,x,y) &&= 0,
\end{eqnarray}
with the computation parameters $\tilde v = 0.5$, $x_0 = 10$, $640 \times 256$ points in the $x \times y$ domain $[0,80] \times [-20, 20]$ and $N_t = 5000$ for $t \in [0,50]$. This yields figure \ref{soliton_quantities} for varying values of $\epsilon$ from $\epsilon = 0.001$ to $\epsilon = 0.1$.

{\color{black} As shown in figure \ref{soliton_quantities}(\textit{a}), the 2D Boussinesq--Peregrine system conserves the mass of the initial condition to machine double precision, and figure \ref{soliton_quantities}(\textit{b})} depicts the  logarithms of the time rates of change of momentum and energy  at $t = 50$ against $\ln(\epsilon)$.  {\color{black} The red and black lines are straight lines obtained by the MATLAB \textit{polyfit} function  and have gradients close to $2$ and $1$, respectively, confirming the \textit{O}$(\epsilon^2)$ and \textit{O}$(\epsilon)$ scaling of the rates of change of momentum and energy (see a relevant discussion in \citealt{DCMM}).  For all numerical simulations in this paper, the mass, momentum and energy are computed for computation parameters such that the quantities listed above are conserved to the precision of the model. }

\section{Numerical Method}\label{appB}

In this section, we discuss the numerical solutions of the KdV$\theta$, KPII, and 2D Boussinesq--Peregrine equations. We use an efficient pseudospectral method where the spatial derivatives are approximated using direct and inverse fast Fourier transforms,  and the temporal derivatives are calculated via a $4^{\text{th}}$-order Runge--Kutta (RK4) scheme for its preferable accuracy.

 For a given discretised field $\eta$ the derivatives are approximated as
\begin{eqnarray}
\textcolor{black}{ \frac{\partial^n \eta}{\partial x^n} = \mathcal{F}^{-1}\left[ (ik_x)^n \mathcal{F}[\eta] \right] \quad \text{and} \quad \frac{\partial^n \eta}{\partial y^n} = \mathcal{F}^{-1}\left[ (ik_y)^n \mathcal{F}[\eta] \right], }
\end{eqnarray}
where $k_x$ and $k_y$ are the wavenumbers, $\mathcal{F}$ is the 2D discrete Fourier transform and $\mathcal{F}^{-1}$ is the 2D inverse discrete Fourier transform. The wavenumbers for computation are matrices of size $N_y \times N_x$ where the rows of $k_x$ are given by
\begin{equation}
0,1,\ldots, \frac{N_x}{2}, -\frac{N_x}{2}+1, -\frac{N_x}{2}+2, \ldots, -1,
\end{equation}
and the columns of $k_y$ by
\begin{equation}
0,1,\ldots, \frac{N_y}{2}, -\frac{N_y}{2}+1, -\frac{N_y}{2}+2, \ldots, -1
\end{equation}
\citep{T}. The Fourier transform of a quantity is denoted below as $\mathcal{F}[\eta] = \hat\eta$. We solve in an arbitrary rectangular domain, meaning we scale the equation from the domain $[-\pi,\pi] \times [-\pi,\pi]$ to $[x_{\text{min}},x_{\text{max}}] \times [y_{\text{min}},y_{\text{max}}]$. Performing this scaling has the effect of scaling $k_x$ and $k_y$ wavenumbers by $2\pi/(x_{\text{max}}-x_{\text{min}})$ and $2\pi/(y_{\text{max}}-y_{\text{min}})$, respectively. The domains are given by
\begin{gather}
x_i = x_{\text{min}} + n \Delta_x, \quad i = 0,1, \ldots, N_x-1, \\
y_j = y_{\text{min}} + m \Delta_y, \quad j = 0,1, \ldots, N_y-1,
\end{gather}
such that $\eta$, $u$ and $v$ are also discretised into $N_y \times N_x$ grids.

The temporal discretisation used is $\Delta_t = (t_{\text{max}}-t_{\text{min}})/N_t$ such that $t_n = t_{\text{min}} + n \Delta_t$, for $n = 0,1,\ldots, N_t$, and is computed via the RK4 scheme such that for a problem in the form
\begin{align}
\eta_t = f(t,\eta),
\end{align}
the next time step, $\eta^{n+1}$, is given as
\begin{align}
\eta^{n+1} = \eta^n + \frac{\Delta_t}{6} \left( k_1 + 2k_2 + 2k_3 + k_4 \right),
\end{align}
where
{\color{black} \begin{align}
k_1 &= f\left( t_n, \eta^n \right), \quad
k_2 = f\left( t_n + \frac{\Delta_t}{2}, \eta^n + \frac{k_1 \Delta_t}{2} \right), \\
k_3 &= f\left( t_n + \frac{\Delta_t}{2}, \eta^n + \frac{k_2 \Delta_t}{2} \right), \quad
k_4 = f\left( t_n + \Delta_t, \eta^n + k_3 \Delta_t \right).
\end{align} }

The 2D Boussinesq--Peregrine equations are solved using the method presented by \citet{ES, SSL} to deal with the two separate time derivatives in (\ref{Bouss2}). This is done by introducing a new variable $q$ such that $q = \nabla \cdot \mathbf{u}_t$. Upon taking the divergence of (\ref{Bouss2}), a third coupled equation is generated, and this gives the following new system of equations with an additional auxiliary equation:
\begin{eqnarray}
&&\eta_t = - \nabla \cdot \{(1 + \epsilon \eta) \mathbf{u} \}, \label{NBouss1}\\
&&\mathbf{u}_t = - \epsilon (\mathbf{u} \cdot \nabla ) \mathbf{u} - \nabla \eta + \frac{\delta^2}{3} \nabla q, \label{NBouss2} \\
&&\frac{\delta^2}{3} \nabla \cdot (\nabla q) - q = \nabla \cdot \left[\epsilon (\mathbf{u} \cdot \nabla) \mathbf{u} + \nabla \eta \right]. \label{qeqn}
\end{eqnarray}
{\color{black} At each time step, the auxiliary equation (\ref{qeqn}) is inverted to find $q$, which can be efficiently done in Fourier space, see \citet{SSL} for details. The data for $q$ is then used in the simultaneous time stepping of (\ref{NBouss1}) and (\ref{NBouss2}). }

The KdV$\theta$ equation has the form
\begin{equation}
\mu_1 A_T + \mu_2 AA_{\xi} + \mu_3 A_{\xi\xi\xi} + \mu_5 \frac{A_{\theta}}{T} = 0.
\end{equation}
Applying the 2D Discrete Fourier transform yields
\begin{equation}
\hat A_T = - \frac{ik_{\xi} \mu_2}{2 \mu_1} \mathcal{F}\left[ A^2 \right] + \frac{ik_{\xi}^3 \mu_3}{\mu_1} \hat A - \frac{ik_{\theta} \mu_5}{\mu_1} \frac{\hat A}{T},
\end{equation}
which is then efficiently iterated in time using the RK4 algorithm discussed above. One must choose the values of $T$ to not include zero  to avoid division by zero, similar to the values at the origin for $R = 0$ for the radial case.

To discretise and numerically solve the KPII equation, we follow the method presented by \citet{KSM,BBH}. To account for both the inward- and outward-propagating versions, we rewrite the KPII equation as
\begin{equation}
\left( 2\eta_T + \frac{3\sigma}{2} \left(\eta^2 \right)_{\xi} + \frac{\sigma}{3}\eta_{\xi\xi\xi}\right)_{\xi} + \sigma \eta_{YY} = 0, \label{appKP}
\end{equation}
where $\sigma$  is $1$ or $-1$  for the right-/left-propagating case, respectively, and $\xi = x - \sigma t$. Applying the 2D Fourier transform to (\ref{appKP}) and rearranging gives
\begin{equation}
\hat\eta_T = - \frac{3\sigma ik_{\xi}}{4} \mathcal{F}\left[\eta^2 \right] + \frac{\sigma ik_{\xi}^3}{6} \hat\eta - \frac{\sigma ik_Y^2}{2k_{\xi}} \hat\eta. \label{discKP}
\end{equation}
It is noted that the form of (\ref{discKP}) includes division by zero for the zero-wavenumber in $k_{\xi}$, however, this relates to the zero-mass condition of the KPII equation. If the zero-mass condition is satisfied at $T_0$, then there is no growth in the zero mode for all $T^n$, and so it is satisfactory to remove the mass from the initial condition and at each time step set $\hat \eta (T^n,0,k_Y) = 0$, see \citet{KSM,BBH} and the references therein for details. Due to the form (\ref{discKP}), it is convenient to introduce the integrating factor
\begin{equation}
\Lambda = \exp \left(-\sigma i \left( \frac{k_{\xi}^3}{6} - \frac{k_Y^2}{2k_{\xi}} \right)T \right), \label{intfac}
\end{equation}
such that $\hat q = \Lambda \hat\eta$. Substituting this expression for $\hat \eta$ into (\ref{discKP}) removes the stiff terms from the problem, and we obtain
\begin{equation}
\textcolor{black}{\hat q_T = - \frac{3 \sigma ik_{\xi} \Lambda}{4} \mathcal{F}\left[ \eta^2 \right],}
\end{equation}
which is a simple ODE for $\hat q$. At each time step, we calculate $\eta^n$ from $\hat q^n = \Lambda \hat\eta^n$ and then calculate $\hat q^{n+1}$. For the KPII equation,  performing additional Fourier transforms to invert the integrating factor is still computationally more efficient than solving the original stiff problem.

\section{Reduction to the KdV equation}\label{appC}

{\color{black}For plane surface and internal waves co-propagating with a shear flow, it can be shown that the amplitude equation (\ref{A2}) reduces to the previously known KdV equation (see, for example, \citealt{G} evaluated for the case of a flat bottom). Indeed, for the waves co-propagating with the current,  the general solution is $k = a \cos \theta$ and  $\xi = rk(\theta) - st = a(x - \hat{s}t)$, where  we have scaled $s \rightarrow a\hat{s}$. The modal equations (\ref{modal1})--(\ref{modal3}) reduce to }
\begin{eqnarray}
\left[ \rho_0 (u_0 - \hat{s})^2 \phi_z \right]_z - \rho_{0z} \phi &&= 0 \quad \text{for} \quad 0 < z < 1, \label{gmodal1} \\
(u_0 - \hat{s})^2 \phi_z  - \phi &&= 0 \quad \text{at} \quad z = 1, \label{gmodal2} \\
\phi &&= 0 \quad \text{at} \quad z = 0. \label{gmodal3}
\end{eqnarray}
Scaling $\xi = rk(\theta) - st = a(x - \hat{s}t)$ to become $\hat{\xi} = x - \hat{s}t$, assuming $A_\theta = 0$ (there is no change in the tangential direction along the wavefront) and computing the remaining coefficients in (\ref{A2}) we obtain
\begin{equation}
I_1 A_T + I_2 AA_{\hat{\xi}} + I_3 A_{\hat{\xi}\hat{\xi}\hat{\xi}} = 0, \label{gKdV}
\end{equation}
where $T = \epsilon t$, $\hat{\xi} = x - \hat{s}t$ and
\begin{eqnarray}
&&I_1 = 2 \int_0^1 \rho_0 (u_0 - \hat{s}) \phi_z^2 ~ \mathrm{d}z, \label{gI1} 
\quad I_2 = - 3 \int_0^1 \rho_0 (u_0 - \hat{s})^2 \phi_z^3 ~ \mathrm{d}z, 
%\label{gI2} 
\\
&&\hspace{2cm} I_3 = - \int_0^1 \rho_0 (u_0 - \hat{s})^2 \phi^2 ~ \mathrm{d}z. \label{gI3}
\end{eqnarray}
The amplitude equation (\ref{gKdV}) and coefficients (\ref{gI1})--(\ref{gI3}) match those of \citet{G} identically. 

{\color{black} The amplitude equation for obliquely propagating plane waves was given by \citet{M}.  This equation can also be recovered from (\ref{A2}) for  $\hat{\xi} = x + b(a)y/a - \hat{s} t$. It has the form (\ref{gKdV}), where the coefficients are given by}
\begin{eqnarray}
&&I_1 = 2 \int_0^1 \rho_0 (u_0 - \hat{s}) \phi_z^2 ~ \mathrm{d}z, \quad
I_2 = - 3 \int_0^1 \rho_0 (u_0 - \hat{s})^2 \phi_z^3 ~ \mathrm{d}z, \\
&&\hspace{2cm} I_3 = - \frac{[a^2 + b(a)^2]}{a^2}\int_0^1 \rho_0 (u_0 - \hat{s})^2 \phi^2 ~ \mathrm{d}z,
\end{eqnarray}
%Once again, for plane waves the term $A_{\theta} = 0$ since there is no change in the tangential direction along the wavefront. 
and the modal equations have the form
\begin{eqnarray}
&&\left[ \frac{\rho_0 a^2 (u_0 - \hat{s})^2}{a^2 + b(a)^2} \phi_z \right]_z - \rho_{0z} \phi = 0 \quad \text{for} \quad 0 < z < 1, \label{obmodal1} \\
&&\frac{a^2(u_0 - \hat{s})^2}{{a^2 + b(a)^2}} \phi_z  - \phi = 0 \quad \text{at} \quad z = 1, \label{obmodal2} \\
&&\phi = 0 \quad \text{at} \quad z = 0. \label{obmodal3}
\end{eqnarray}
{\color{black} The modal equations (\ref{obmodal1})--(\ref{obmodal3}) reduce to (\ref{gmodal1})--(\ref{gmodal3}) when the wave is co-propagating with the current, since in this case $b(a) = 0$.}

% ---------- ---------- ---------- ---------- %
\bibliographystyle{jfm}

\begin{thebibliography}{99}

\expandafter\ifx\csname natexlab\endcsname\relax
\def\natexlab#1{#1}\fi
\expandafter\ifx\csname selectlanguage\endcsname\relax
\def\selectlanguage#1{\relax}\fi

\bibitem[Ablowitz \& Baldwin (2012)]{AB}
{\sc Ablowitz, M. \& Baldwin, D.} 2012 {Nonlinear shallow ocean-wave soliton interactions on flat beaches}, {\it Phys. Rev. E}, {\bf 86}, 036305.

\bibitem[Ablowitz \& Curtis (2013)]{AC}
{\sc Ablowitz, M. \& Curtis, S.} 2013 {Conservation laws and non-decaying solutions for the Benney–Luke equation}, {\it Proc. R. Soc. A}, {\bf 469}, 20120690.

\bibitem[Ablowitz \& Segur (1979)]{AS}
{\color{black}{\sc Ablowitz, M. \& Segur, H.} 1978 {On the evolution of packets of water waves}, {\it J. Fluid Mech.}, {\bf 92}, pp. 691-715.}

\bibitem[Apel (2003)]{A}
{\sc Apel, J.} 2003 {A new analytical model for internal solitons in the ocean}, {\it J. Phys. Oceanogr.}, {\bf 33}, pp. 2247-2269.

\bibitem[Apel {\it et al.} (2007)]{AOSL} 
{\color{black}{\sc Apel, J., Ostrovsky, L., Stepanyants, Y. \& Lynch, J.} 2007 Internal solitons in the ocean and their effect on underwater sound, {\it J. Acoust. Soc. Am.}, {\bf 121}, pp. 695 - 722.}

\bibitem[Barros \& Choi (2014)]{BC}
{\sc Barros, R. \& Choi, W.} 2014 {Elementary stratified flows with stability at low Richardson Number}, {\it Phys. Fluids}, {\bf 26}, 124107.

\bibitem[Bassi {\it et al.} (2020)]{BBBD}
{\sc Bassi, C., Bonaventura, L., Busto, S. \& Dumbser, M.} 2020 {A hyperbolic reformulation of the Serre--Green--Naghdi model for general bottom topographies}, {\it Computers \& Fluids}, {\bf 212}, 104716.

%\bibitem[Benjamin (1966)]{B1}  Benjamin, T.B. 1966 Internal waves of finite amplitude and permanent form. {\it J. Fluid Mech.} {\bf 25}, 241 - 270.

%\bibitem[Benney (1966)]{B2} Benney, D.J. 1966 Long nonlinear waves in fluid flows. {\it J. Math. Phys.} {\bf 45}, 52 - 63.

\bibitem[Biondini, Bivolcic \& Hoefer (2025)]{BBH}
{\sc Biondini, B., Bivolcic, A. \& Hoefer, M.} 2025 {Mach reflection and expansion of two-dimensional dispersive shock waves}, {\it Phys. Rev. Lett.}, {\bf 135}, 067201.

\bibitem[Biondini, Hoefer \& Moro (2020)]{BHM}
{\sc Biondini, G., Hoefer, M. \& Moro, A.} 2020 {Integrability, exact reductions and special solutions of the KP–Whitham equations}, {\it Nonlinearity}, {\bf 33}, pp. 4114-4132.

\bibitem[Bona et al. (2008)]{BLS}
{\color{black}{\sc Bona, J., Lannes, D. \& Saut, J.} 2008 {Asymptotic models for internal waves}, {\it J. Math. Pure Appl.}, {\bf 89}, pp. 538-566.}

\bibitem[Boonkasame \& Milewski (2011)]{BM} 
{\sc Boonkasame, A. \& Milewski, P.} 2011 {The stability of large-amplitude shallow interfacial non-Boussinesq flows}, {\it Stud. Appl. Math.}, {\bf 133}, pp. 182-213.

%\bibitem[Boussinesq (1871)]{Bo} Boussinesq, J. 1871 Th\'eorie de l'intumescence liquide appel\'ee onde solitaire ou de translation, se propageant dans un canal rectangulaire. {\it Comptes Rendus Acad. Sci. (Paris)} {\bf 72}, 755 - 759.

\bibitem[Burns (1953)]{B} 
{\sc Burns, J.C.} 1953  {Long waves in running water}, {\it Proc. Camb. Phil. Soc.}, {\bf 49}, pp. 695-706.

\bibitem[Chakravarty \& Kodama (2009)]{CK09}
{\sc Chakravarty, S. \& Kodama, Y.} 2009 {Soliton solutions of the KP equation and application to shallow water waves}, {\it Stud. Appl. Math.}, {\bf 123}, pp. 83-151.

\bibitem[Chakravarty \& Kodama (2014)]{CK}
{\sc Chakravarty, S. \& Kodama, Y.} 2014 {Construction of KP solitons from wave patterns}, {\it J. Phys. A}, {\bf 47}, 025201.

\bibitem[Chakravarty, Lewkow \& Maruno (2010)]{CLM}
{\sc Chakravarty, S., Lewkow, T. \& Maruno, K.} 2010 {On the construction of the KP line-solitons and their interactions}, {\it Applicable Analysis}, {\bf 89}, pp. 529-545.

\bibitem[Chwang \& Wu (1977)]{CW}
{\sc Chwang, A. \& Wu, T.} 1977 {Cylindrical solitary waves}, Waves on water of variable depth, {\it Springer}.

\bibitem[Drazin (1983)]{D}
{\sc Drazin, P.} 1983, {Solitons}, London Mathematical Society Lecture Note Series, {\it Cambridge University Press}.

\bibitem[Dutykh et al. (2013)]{DCMM}
{\color{black} {\sc Durykh, D., Clamind, D., Milewski, P. \& Mitsotakis, D.} 2013 {Finite volume and pseudospectral schemes for the fully nonlinear 1D Serre equations}, {\it Eur. J. Appl. Math.}, {\bf 24}, pp. 761-787.}

\bibitem[Ellingsen (2014a)]{E}
{\sc Ellingsen, S.} 2014a {Ship waves in the presence of uniform vorticity}, {\it J. Fluid Mech.}, {\bf 742}, R2.

\bibitem[Eskilsson \& Sherwin (2006)]{ES}
{\sc Eskilsson, C.  \& Sherwin, S.} 2006 {Spectral/hp discontinuous Galerkin methods for modelling 2D Boussinesq equations}, {\it J. Comput. Phys.}, {\bf 212}, pp. 566-589.

\bibitem[Gardner  {\it et al.} (1967)]{GGKM} 
{\sc Gardner, C., Green, J., Kruskal, M. \& Miura R.} 1967 {Method for solving the Korteweg--de Vries equation}, {\it Phys. Rev. Lett.},  {\bf 19}, pp. 1095-1097. 

\bibitem[Gavrilyuk \& Klein (2024)]{GK}
{\sc Gavrilyuk, S. \& Klein, C.} 2024 {Numerical study of the Serre--Green--Naghdi equations in 2D}, {\it Nonlinearity}, {\bf 37}, 045014.

\bibitem[Gavrilyuk \& Shyue (2024)]{GS}
{\sc Gavrilyuk, S. \& Shyue, K.} 2024 {2D Serre--Green--Naghdi equations over topography: elliptic operator inversion method}, {\it J. Hydraulic Eng.}, {\bf 150}, 04023054.

\bibitem[Grimshaw (2005)]{G}
{\sc Grimshaw, R.} 2005 {Nonlinear Waves in Fluids: Recent Advances and Modern Applications}, {\it Springer.}

%\bibitem[Grimshaw {\it et al.} (1997)]{GPT} Grimshaw, R.H.J., Pelinovsky, E. $\&$ Talipova, T. 1997 The modified {Korteweg - de Vries} equation in the theory of large-amplitude internal waves. {\it Nonlin. Processes Geophys.} {\bf 4}, 237 - 250.

\bibitem[Grimshaw {\it et al.} (1998)]{GOSS}  
{\color{black}{\sc Grimshaw, R., Ostrovsky, L., Shrira, V. \& Stepanyants, Y.} 1998 {Long nonlinear surface and internal gravity waves in a rotating ocean}, {\it Surveys in Geophysics}, {\bf 19}, pp. 289-338.}

\bibitem[Grimshaw et al. (2010)]{GPTK}
{\color{black}{\sc Grimshaw R., Pelinovsky E., Talipova T. \& Kurkina O.} 2010 {Internal solitary waves: propagation, deformation and disintegration}, {\it Nonlin Proc. Geophys.}, {\bf 17}, pp. 633-649.}

\bibitem[Grimshaw et al. (2013)]{GHJ}
{\color{black}{\sc Grimshaw R., Helfrich K. \& Johnson E.} 2013. {Experimental study of the effect of rotation on large amplitude internal waves}, {\it Phys Fluids.}, {\bf 25}, 056602.}

\bibitem[Grue (2006)]{Gr}  
{\color{black}{\sc Grue, J.} 2006 {Very large internal waves in the ocean - observations and nonlinear models}, {\it Waves in Geophysical Fluids} (ed. J. Grue \& K. Trulsen), pp. 1 - 66. Springer.}

\bibitem[Haragus, Li \& Pelinovsky (2017)]{HP}
{\sc Haragus, M., Li, J. \& Pelinovsky, D.} 2017 {Counting unstable eigenvalues in Hamiltonian spectral problems via commuting operators}, {\it Comm. Math. Phys.}, {\bf 354}, pp. 247-268.

\bibitem[He \& Chabchoub (2025)]{HC}
{\sc He, Y. \& Chabchoub, A.} 2025 {On long-crested ocean rogue waves originating from localized amplitude and frequency modulations}, {\it Ocean Modelling}, {\bf 193}, 102464.

\bibitem[Helfrich \& Melville (2006)]{HM} 
{\color{black}{\sc Helfrich, K. \& Melville, W.} 2006 {Long nonlinear internal waves}, {\it Ann. Rev. Fluid Mech.}, {\bf 38}, pp. 395-425.}

\bibitem[Hooper, Khusnutdinova \& Grimshaw (2021)]{HKG} 
{\sc Hooper, C., Khusnutdinova, K. \& Grimshaw, R.} 2021 {Wavefronts and modal structure of long surface and internal ring waves on a parallel shear current}, {\it J. Fluid Mech.}, {\bf 927}, A37.

\bibitem[Hornick, Pelinovsky \& Schneider (2025)]{HPS}
{\sc Hornick, J., Pelinovsky, D. \& Schneider, G.} 2025 {On the long-wave approximation of solitary waves in cylindrical coordinates}, {\it Nonlinear Differ. Equ. Appl.}, {\bf 32}, 50.

\bibitem[Hu, Luo \& Wang (2025)]{HLW}
{\sc Hu, L., Luo, X. \& Wang, Z.} 2025 {Obliquely interacting solitary waves and wave wakes in free-surface flows}, {\it J. Fluid Mech.}, {\bf 1011}, A8.

\bibitem[Johnson (1990)]{J90}
{\sc Johnson, R.} 1990 {Ring waves on the surface of shear flows: a linear and nonlinear theory}, {\it J. Fluid Mech.}, {\bf 215}, pp. 145-160.

\bibitem[Johnson (1997)]{J97}
{\sc Johnson, R.} 1997 {A modern introduction to the mathematical theory of water waves}. Cambridge University Press.

\bibitem[Johnson (1999)]{J99} 
{\sc Johnson, R.} 1999 {A note on an asymptotic solution of the cylindrical Korteweg-de Vries equation}, {\it Wave Motion}, {\bf 30}, pp. 1-16.

\bibitem[Kadomtsev \& Petviashvili (1970)]{KP}
{\sc Kadomtsev, B. \& Petviashvili, V.} 1970 {On the Stability of Solitary Waves in Weakly Dispersing Media}, {\it Soviet Physics Doklady}, {\bf 15}, pp. 539-541.

\bibitem[Katsaounis, Mitsotakis \& Sadaka (2020)]{KMS}
{\sc Katsaounis, T., Mitsotakis, D. \& Sadaka, G.} 2020 {Boussinesq--Peregrine water wave models and their numerical approximation}, {\it J. Comput. Phys.}, {\bf 417}, 109579.

\bibitem[Kharif, Pelinovsky \& Slunyaev (2008)]{KPS}
{\sc Kharif, C., Pelinovsky, E. \& Slunyaev, A.} 2008 {Rogue waves in the ocean}, {\it Advances in Geophysical and Environmental Mechanics and Mathematics}, Springer.

\bibitem[Khorbatly \& Kalisch (2024)]{KK}
{\sc Khorbatly, B. \& Kalisch, H.} 2024 {Rigorous estimates on mechanical balance laws in the Boussinesq–Peregrine equations}, {\it Stud. Appl. Math.}, {\bf 152}, pp. 847-867.

\bibitem[Khusnutdinova (2020)]{K}
{\sc Khusnutdinova, K.} 2020 {Long internal ring waves in a two-layer fluid with an upper-layer current}, {\it Russ. J. Earth Sci.}, {\bf 20}, ES4006.

\bibitem[Khusnutdinova \& Zhang (2016)]{KZ}
{\sc Khusnutdinova, K. \& Zhang, X.} 2016a {Long ring waves in a stratified fluid over a shear flow}, {\it J. Fluid Mech.}, {\bf 794}, pp. 17-44.

\bibitem[Kivshar \& Pelinovsky (2000)]{KP00}
{\sc Kivshar, Y. \& Pelinovsky, D.} 2000 {Self-focusing and transverse instabilities of solitary waves}, {\it Phys. Rep.}, {\bf 331}, pp. 117-195.

\bibitem[Klein, Sparber \& Markowich (2007)]{KSM}
{\sc Klein, C., Sparber, C. \& Markowich, P.} 2007 {Numerical Study of Oscillatory Regimes in the Kadomtsev–Petviashvili Equation}, {\it J. Nonlinear Sci.}, {\bf 17}, pp. 429-470.

%\bibitem[Korteweg $\&$ de Vries (1895)]{KdV} Korteweg, D.J. $\&$ de Vries, G. 1895 On the change of form of long waves advancing in a rectangular channel, and on a new type of long stationary waves. {\it Philos. Mag.} {\bf 39}, 422 - 443.

%\bibitem[Ko \& Kuel (1979)]{KoK}
%{\color{black}{\sc Ko, K., Kuel, H.H.} 1979 {Cylindrical and spherical KdV solitary waves}, {\it Phys. Fluids}, {\bf 22}, pp. 1343-1348.}

\bibitem[Ko \& Kuel (1979)]{KoK}
{\color{black}{\sc Ko, K. \& Kuel, H.} 1979 {Cylindrical and spherical KdV solitary waves}, {\it Phys. Fluids}, {\bf 22}, pp. 1343-1348.}

\bibitem[Dorfman, Pelinovsky \& Stepanyants (1981)]{DPS}
{\color{black}{\sc Dorfman, A., Pelinovsky, E. \& Stepanyants, Y.} 1981 {Finite-amplitude cylindrical and spherical waves in weakly dispersive media}, {\it Sov. Phys. J. Appl. Mech. Tech. Phys.}, {\bf 2}, pp. 206-211.}

\bibitem[Krechetnikov (2024)]{K24}
{\sc Krechetnikov, R.} 2024 {Transverse instability of concentric water waves}, {\it J. Nonlinear Sci.}, {\bf 34}, 66.

\bibitem[Lamb (2005)]{L}
{\color{black}{\sc Lamb, K.} 2005 {Extreme internal solitary waves in the ocean: theoretical considerations}, Aha Huliko’a Hawaiian Winter Workshop.}

\bibitem[Landau \& Lifshitz (1959)]{LL}
{\sc Landau, L. \& Lifshitz, E.} 1959 {Quantum Mechanics}, {\it Pergamon Press}.

\bibitem[Lannes \& Ming (2015)]{LM}
{\sc Lannes, D. \& Ming, M.} 2015 {The Kelvin-Helmholtz instabilities in two-fluids shallow water models}, {\it Hamiltonian Partial Differential Equations and Applications}, Fields Institute Comms, vol. 75, Springer.

\bibitem[Li \& Ellingsen (2019)]{LE}
{\sc Li, Y. \&  Ellingsen, S.} 2019  {A framework for modelling linear surface waves on shear currents in slowly varying waves}, {\it J. Geophys. Res.: Oceans}, {\bf 124}, pp. 2527-2545.

%\bibitem[Maslowe $\&$ Redekopp (1980)]{MR} Maslowe, S.A. $\&$ Redekopp, L.G. Long nonlinear waves in stratified shear flows. {\it J. Fluid Mech.} {\bf 101}, 321 - 348.

\bibitem[Miles (1977)]{M}
{\sc Miles, J.} 1977 {Obliquely interacting solitary waves}, {\it J. Fluid Mech.}, {\bf 79}, pp. 157-169.

\bibitem[Miles (1978)]{M2}
{\color{black}{\sc Miles, J.} 1978 {An axisymmetric Boussinesq wave}, {\it J. Fluid Mech.}, {\bf 84}, pp. 181-191.}

\bibitem[Mizumachi (2015)]{Mi}
{\sc Mizumachi, T.} 2015 {Stability of line solitons for the KP-II equation in R2}, {\it Mem. Am. Math. Soc.}, {\bf 238}, 1125.

\bibitem[Mizumachi \& Shimabukuro (2017)]{MS1}
{\sc Mizumachi, T. \& Shimabukuro, Y.} 2017 {Asymptotic linear stability of Benney-Luke line solitary waves in 2D}, {\it Nonlinearity}, {\bf 30}, pp. 3419-3465.

\bibitem[Mizumachi \& Shimabukuro (2020)]{MS2}
{\sc Mizumachi, T. \& Shimabukuro, Y.} 2020 {Stability of Benney-Luke line solitary waves in 2 dimensions}, 
{\it SIAM J. Math. Anal.}, {\bf 52}, pp. 4238-4283.

\bibitem[Mizumachi \& Tzvetkov (2012)]{MT}
{\sc Mizumachi, T. \& Tzvetkov, N.} 2012 {Stability of the line soliton of the KP-II equation under periodic transverse perturbations}, {\it Math. Ann.}, {\bf 352}, pp. 659-690.

\bibitem[Nirunwiroj, Tseluiko \& Khusnutdinova (2025)]{NTK}
{\color{black} {\sc Nirunwiroj, K., Tseluiko, D. \& Khusnutdinova, K.} {Evolution of internal cnoidal waves with local
defects in a two-layer fluid with and without rotation}, {\it J. Fluid Mech.} (2025) accepted.}

\bibitem[Onorato {\it et al.} (2013)]{ORBMA}
{\sc Onorato, M., Residori, S., Bortolozzo, U., Montina, A., \& Arecchi, F.} 2013 {Rogue waves and their generating mechanisms in different physical contexts}, {\it Phys. Rep.}, {\bf 528}, pp. 47-89.

\bibitem[Ostrovskii \& Shrira (1976)]{OS} 
{\sc Ostrovskii, L. \& Shrira, V.} 1976 {Instability and self-refraction of solitons}, {\it Sov. J. Exp. Theor. Phys.}, {\bf  44}, pp. 738-743.

\bibitem[Ostrovsky {\it et al.} (2015)]{OPSS}
{\color{black}{\sc Ostrovskii, L., Pelinovsky, E., Shrira, V. \& Stepanyants, Y.} 2015 {Beyond the KdV: Post-explosion development}, {\it Chaos}, {\bf 25}, 097620.}

\bibitem[Ovsyannikov (1979)]{O}
{\sc Ovsyannikov, L.} 1979 {Two-layer ``shallow-water'' model}, {\it J. Appl. Math. Tech. Phys.}, {\bf 20}, pp. 127-135.

\bibitem[Peregrine (1967)]{P}
{\sc Peregrine, D.} 1967 {Long waves on a beach}, {\it J. Fluid Mech.}, {\bf 27}, pp. 815-827.

\bibitem[Pesenson (1991)]{P91}
{\sc Pesenson, M.} 1991 {Nonlinear waves travelling upon a front of solitons}, {\it Phys. Fluids}, {\bf 3}, pp. 3001-3006.

\bibitem[Porubov {\it et al.} (2005)]{PTLO} 
{\color{black} {\sc Porubov, A., Truji, H., Lavrenov, I. \& Oikawa, M.} 2005 {Formation of the rogue wave due to non-linear two-dimensional waves interaction}, {\it Wave Motion}, {\bf 42}, pp. 202-210.}

\bibitem[Ryskamp {\it et al.} (2021)]{RMBH}
{\sc Ryskamp, S., Maiden, M., Biondini, G. \& Hoefer, M.} 2021 {Evolution of truncated and bent gravity wave solitons: the Mach expansion problem}, {\it J. Fluid Mech.}, {\bf 909}, A24.

\bibitem[Ryskamp, Hoefer \& Biondini (2022)]{RHB}
{\sc Ryskamp, S., Hoefer M. \& Biondini, G.} 2022 {Modulation theory for soliton resonance and Mach reflection}, {\it Proc. R. Soc. A}, {\bf 478},  20210823.

\bibitem[Shrira \& Pesenson (1984)]{SP}
{\sc Shrira, V. \& Pesenson, M.} 1984 {Soliton stability with respect to transverse perturbations of finite length and amplitude}, {\it Nonlinear and Turbulent Processes in Physics}, Harwood Acad. Publ., pp. 1061-1068.

\bibitem[Sidorovas {\it et al.} (2024)]{STCK}
{\sc Sidorovas, N., Tseluiko, D., Choi, W. \& Khusnutdinova, K.} 2024 {Nonlinear concentric water waves of moderate amplitude}, {\it Wave Motion}, {\bf 128}, 103295.

\bibitem[Sidorovas {\it et al.} (2025)]{STCK2}
{\color{black}{\sc Sidorovas, N., Tseluiko, D., Choi, W. \& Khusnutdinova, K.} 2025 {Internal solitary and cnoidal waves of moderate amplitude in a two-layer fluid: the extended KdV equation approximation}, {\it Phys. D}, {\bf 481}, 134723.}

\bibitem[Slunyaev \& Pelinovsky (2016)]{SP16}
{\sc Slunyaev, A. \& Pelinovsky, E.} 2016 {Role of multiple soliton interactions in the generation of rogue waves: the modified Korteweg-de Vries framework}, {\it Phys. Rev. Lett.}, {\bf 117}, 214501.

\bibitem[Smeltzer, Æsøy \& Ellingsen (2019)]{SEE} 
{\sc Smeltzer, B., Æsøy, E. \&  Ellingsen, S.} 2019 {Observation of surface wave patterns modified by sub-surface shear currents}, {\it J. Fluid Mech.}, {\bf 873}, pp. 508-530.

\bibitem[Stastna (2022)]{S}
{\color{black}{\sc Stastna, M.} 2022 {Internal waves in the ocean: theory and practice}, {\it Springer}.}

\bibitem[Steinmoeller, Stastna \& Lamb (2012)]{SSL}
{\sc Steinmoeller, D., Stastna, M. \& Lamb, K.} 2012 {Fourier pseudospectral methods for 2D Boussinesq-type equations}, {\it Ocean Modelling}, {\bf 52}, pp. 76-89.

\bibitem[Trefethen (2000)]{T}
{\sc Trefethen, L.} 2000 {Spectral Methods in MATLAB}, Software, environments, tools, {\it SIAM}.

\bibitem[Tseluiko {\it et al.} (2023)]{TABK} 
{\sc Tseluiko, D., Alharthi, N., Barros, R. \& Khusnutdinova, K.} 2023 {Internal ring waves in a three-layer fluid on a current with a constant vertical shear}, {\it Nonlinearity}, {\bf 36}, pp. 3431-3466.

\bibitem[Yuan {\it et al.} (2018a)]{YGJW}
{\sc Yuan, C., Grimshaw, R., Johnson, E. \& Wang, Z.} 2018a {Topographic effect on oblique internal wave–wave interactions}, {\it J. Fluid Mech.}, {\bf 836}, pp. 36-60.

\bibitem[Yuan \& Wang (2022)]{YW}
{\sc Yuan, C. \& Wang, Z.} 2022 {On diffraction and oblique interactions of horizontally two-dimensional internal solitary waves}, {\it J. Fluid Mech.}, {\bf 936}, A20.

\bibitem[Zhang {\it et al.} (2024)]{ZHGS}
{\sc Zhang, Z., Hu, W., Guo, Q. \& Stepanyants, Y.} 2024 {Solitons and lumps in the cylindrical Kadomtsev–Petviashvili equation. II. Lumps and their interactions}, {\it Chaos}, {\bf 34}, 013132.

\end{thebibliography}

\end{document}